\documentclass[twocolumn]{aastex62}
\usepackage{amsmath}
\usepackage{graphicx}
\usepackage{mathtools}
\usepackage{multirow}

\usepackage{xcolor}

\newcommand{\HI}{\mbox{H \textsc{I}}}
\newcommand{\HII}{\mbox{H \textsc{II}}}
\newcommand{\OIII}{\mbox{[O \textsc{III}]}}
\newcommand{\OII}{\mbox{[O \textsc{II}]}}
\newcommand{\NII}{\mbox{[N \textsc{II}]}}
\newcommand{\SII}{\mbox{[S \textsc{II}]}}

\newcommand{\Ha}{H$\alpha$}
\newcommand{\Hb}{H$\beta$}
\newcommand{\OIIIHb}{\OIII/H$\beta$}
\newcommand{\NIIHa}{\NII/H$\alpha$}
\newcommand{\SIIHa}{\SII/H$\alpha$}
\newcommand{\SIIb}{\SII$\lambda$6716}
\newcommand{\SIIr}{\SII$\lambda$6731}
\newcommand{\OIIb}{\OII$\lambda$3726}
\newcommand{\OIIr}{\OII$\lambda$3729}
\newcommand{\ktd}{KMOS$^{\rm 3D}$}
\newcommand{\reff}{$R_e$}
\newcommand{\sfrsd}{$\Sigma_{\rm SFR}$}
\newcommand{\smsd}{$\Sigma_*$}
\newcommand{\gmsd}{$\Sigma_{H_2}$}
\newcommand{\kms}{\mbox{km s$^{-1}$}}
\newcommand{\cmcubed}{cm$^{-3}$}
\newcommand{\rSII}{$R_{\rm SII}$}
\newcommand{\logpk}{$\log(P_{\rm th}/k)$}

\newcommand{\highz}{\mbox{high-$z$}}

\newcommand{\appropto}{\mathrel{\vcenter{
  \offinterlineskip\halign{\hfil$##$\cr
    \propto\cr\noalign{\kern2pt}\sim\cr\noalign{\kern-2pt}}}}}

\newenvironment{nscenter}
 {\parskip=0pt\par\nopagebreak\centering}
 {\par\noindent\ignorespacesafterend}

\shorttitle{Redshift Evolution of the Electron Density}
\shortauthors{Rebecca L. Davies et al.}

\begin{document}

\title{The \ktd\ Survey: Investigating the Origin of the Elevated Electron Densities in Star-Forming Galaxies at \mbox{1 $\lesssim z \lesssim$ 3}}

\correspondingauthor{Rebecca L. Davies}
\email{rdavies@swin.edu.au}

\author{Rebecca L. Davies}
\affiliation{Max-Planck-Institut f\"ur extraterrestrische Physik,
             Giessenbachstrasse, D-85748 Garching, Germany}
\affiliation{Centre for Astrophysics and Supercomputing, Swinburne University of Technology, Hawthorn, Victoria 3122, Australia}
\affiliation{ARC Centre of Excellence for All Sky Astrophysics in 3 Dimensions (ASTRO 3D), Australia}

\author{N.M. F\"orster Schreiber}
\affiliation{Max-Planck-Institut f\"ur extraterrestrische Physik,
             Giessenbachstrasse, D-85748 Garching, Germany}

\author{R. Genzel}
\affiliation{Max-Planck-Institut f\"ur extraterrestrische Physik,
             Giessenbachstrasse, D-85748 Garching, Germany}             

\author{T.T. Shimizu}
\affiliation{Max-Planck-Institut f\"ur extraterrestrische Physik,
             Giessenbachstrasse, D-85748 Garching, Germany}

\author{R.I. Davies}
\affiliation{Max-Planck-Institut f\"ur extraterrestrische Physik,
             Giessenbachstrasse, D-85748 Garching, Germany}
             
\author{A. Schruba}
\affiliation{Max-Planck-Institut f\"ur extraterrestrische Physik,
             Giessenbachstrasse, D-85748 Garching, Germany}

\author{L.J. Tacconi}
\affiliation{Max-Planck-Institut f\"ur extraterrestrische Physik,
             Giessenbachstrasse, D-85748 Garching, Germany}
             
\author{H. \"Ubler}
\affiliation{Max-Planck-Institut f\"ur extraterrestrische Physik,
             Giessenbachstrasse, D-85748 Garching, Germany}

\author{E. Wisnioski}
\affiliation{Research School of Astronomy \& Astrophysics, Australian National University, Canberra, ACT 2611, Australia}
\affiliation{ARC Centre of Excellence for All Sky Astrophysics in 3 Dimensions (ASTRO 3D), Australia}

\author{S. Wuyts}
\affiliation{Department of Physics, University of Bath, Claverton Down, Bath, BA2 7AY, UK}

\author{M. Fossati}
\affiliation{Dipartimento di Fisica G. Occhialini, Universit\`a degli Studi di Milano-Bicocca, Piazza della Scienza 3, 20126 Milano, Italy}

\author{R. Herrera-Camus}
\affiliation{Departamento de Astronomía, Universidad de Concepción, Barrio Universitario, Concepción, Chile}

\author{D. Lutz}
\affiliation{Max-Planck-Institut f\"ur extraterrestrische Physik,
             Giessenbachstrasse, D-85748 Garching, Germany}

\author{J.T. Mendel}
\affiliation{Research School of Astronomy \& Astrophysics, Australian National University, Canberra, ACT 2611, Australia}
\affiliation{ARC Centre of Excellence for All Sky Astrophysics in 3 Dimensions (ASTRO 3D), Australia}

\author{T. Naab}
\affiliation{Max-Planck-Institut f\"ur Astrophysik, 
	    Karl-Schwarzschildstr. 1, D-85748 Garching, Germany}
             
\author{S.H. Price}
\affiliation{Max-Planck-Institut f\"ur extraterrestrische Physik,
             Giessenbachstrasse, D-85748 Garching, Germany}

\author{A. Renzini}
\affiliation{INAF – Osservatorio Astronomico di Padova, Vicolo dell’Osservatorio 5, I-35122 Padova, Italy}

\author{D. Wilman}
\affiliation{Max-Planck-Institut f\"ur extraterrestrische Physik,
             Giessenbachstrasse, D-85748 Garching, Germany}
\affiliation{Universit\"ats-Sternwarte Ludwig-Maximilians-Universit\"at M\"unchen, Scheinerstr. 1, D-81679 München, Germany}

\author{A. Beifiori}
\affiliation{Max-Planck-Institut f\"ur extraterrestrische Physik,
             Giessenbachstrasse, D-85748 Garching, Germany}
\affiliation{Universit\"ats-Sternwarte Ludwig-Maximilians-Universit\"at M\"unchen, Scheinerstr. 1, D-81679 München, Germany}

\author{S. Belli}
\affiliation{Harvard-Smithsonian Center for Astrophysics, 60 Garden Street, Cambridge, MA 02138, USA}

\author{A. Burkert}
\affiliation{Max-Planck-Institut f\"ur extraterrestrische Physik,
             Giessenbachstrasse, D-85748 Garching, Germany}
\affiliation{Universit\"ats-Sternwarte Ludwig-Maximilians-Universit\"at M\"unchen, Scheinerstr. 1, D-81679 München, Germany}

\author{J. Chan}
\affiliation{Department of Physics and Astronomy, University of California, Riverside, CA 92521, USA}

\author{A. Contursi}
\affiliation{Max-Planck-Institut f\"ur extraterrestrische Physik,
            Giessenbachstrasse, D-85748 Garching, Germany}
\affiliation{IRAM, 300 Rue de la Piscine, 38406 Saint Martin D’Hères, Grenoble, France}

\author{M. Fabricius}
\affiliation{Max-Planck-Institut f\"ur extraterrestrische Physik,
            Giessenbachstrasse, D-85748 Garching, Germany}

\author{M.M. Lee}
\affiliation{Max-Planck-Institut f\"ur extraterrestrische Physik,
             Giessenbachstrasse, D-85748 Garching, Germany}

\author{R.P. Saglia}
\affiliation{Max-Planck-Institut f\"ur extraterrestrische Physik,
             Giessenbachstrasse, D-85748 Garching, Germany}
\affiliation{Universit\"ats-Sternwarte Ludwig-Maximilians-Universit\"at M\"unchen, Scheinerstr. 1, D-81679 München, Germany}

\author{A. Sternberg}
\affiliation{School of Physics \& Astronomy, Tel Aviv University, Ramat Aviv 69978, Israel}
\affiliation{Center for Computational Astrophysics, Flatiron Institute, 162 5th Avenue, New York, NY
10010, USA}             

\begin{abstract}
We investigate what drives the redshift evolution of the typical electron density ($n_e$) in star-forming galaxies, using a sample of 140 galaxies drawn primarily from \ktd\ \mbox{($0.6<z<2.6$)} and 471 galaxies from SAMI \mbox{($z<0.113$)}. We select galaxies that do not show evidence of AGN activity or outflows, to constrain the average conditions within \HII\ regions. Measurements of the \SII$\lambda$6716/\SII$\lambda$6731 ratio in four redshift bins indicate that the local $n_e$ in the line-emitting material decreases from \mbox{187$^{+140}_{-132}$~\cmcubed} at $z\sim{2.2}$ to \mbox{32$^{+4}_{-9}$~\cmcubed} at $z\sim{0}$; consistent with previous results. We use the \Ha\ luminosity to estimate the root-mean-square (rms) $n_e$ averaged over the volumes of star-forming disks at each redshift. The local and volume-averaged $n_e$ evolve at similar rates, hinting that the volume filling factor of the line-emitting gas may be approximately constant across \mbox{$0\lesssim{z}\lesssim{2.6}$}. The \ktd\ and SAMI galaxies follow a roughly monotonic trend between $n_e$ and star formation rate, but the \ktd\ galaxies have systematically higher $n_e$ than the SAMI galaxies at fixed offset from the star-forming main sequence, suggesting a link between the $n_e$ evolution and the evolving main sequence normalization. We quantitatively test potential drivers of the density evolution and find that \mbox{$n_e$(rms) $\simeq{n_{H_2}}$}, suggesting that the elevated $n_e$ in \highz\ \HII\ regions could plausibly be the direct result of higher densities in the parent molecular clouds. There is also tentative evidence that $n_e$ could be influenced by the balance between stellar feedback, which drives the expansion of \HII\ regions, and the ambient pressure, which resists their expansion.
\end{abstract}


\section{Introduction}
The average properties of star-forming galaxies (SFGs) have evolved significantly from the peak epoch of star formation to the present day universe. The cosmic star formation rate (SFR) density and the normalization of the star-forming main sequence (MS) have both decreased by an order of magnitude since $z\sim$~2 \citep[e.g.][]{Daddi07, Elbaz07, Madau14, Sobral14, Speagle14, Whitaker14}, primarily driven by the declining rate of cosmological cold gas accretion and the subsequent reduction in the molecular gas fractions of galaxies \citep[e.g.][]{Genzel15, Scoville17, Liu19, Millard20, Tacconi20}. The high gas fractions at $z\sim$~2 drive galaxy-wide gravitational instabilities, resulting in elevated gas velocity dispersions \citep[e.g.][]{Genzel06, Genzel08, Law09, Newman13, Wisnioski15, Johnson18, Krumholz18, Uebler19} and triggering the formation of massive star-forming clumps \citep[e.g.][]{Elmegreen05, Bournaud07, Dekel09, Genzel11, Genel12, Wisnioski12, Wuyts12}.

It is then perhaps not surprising that we also observe significant evolution of the properties of the interstellar medium (ISM). The first near-infrared spectroscopic surveys of high-redshift SFGs revealed that they do not lie along the locus of local SFGs on the \NIIHa\ vs. \OIIIHb\ diagnostic diagram, but are offset to higher line ratios \citep[e.g.][]{Shapley05, Erb06a, Kriek07}. The physical origin of this offset remains highly debated, with proposed explanations including a harder ionizing radiation field \citep[e.g.][]{Steidel16, Strom17, Strom18, Sanders20}, higher N/O abundance ratio \citep[e.g.][]{Masters14, Masters16, Jones15, Shapley15}, elevated electron density and ISM pressure \citep[e.g.][]{Dopita16, dAgostino19}, higher ionization parameter \citep[e.g.][]{Kashino17, Bian20}, an increased contribution from shocks and/or Active Galactic Nuclei \citep[AGN; e.g.][]{Newman14, Freeman19}, and/or a decreased contribution from diffuse ionized gas within the regions sampled by the observations \citep[e.g.][]{Shapley19}. It is very difficult to distinguish between different possible drivers based on the \NIIHa\ and \OIIIHb\ ratios alone \citep[e.g.][]{Ke13}, and it is necessary to quantify the evolution of each property in order to build a full picture of how the physical conditions in star-forming regions have evolved over time.

\begin{sloppypar}
The electron density is typically measured using density-sensitive line ratios such as \SIIb/\SIIr, \OIIr/\OIIb\ and \mbox{C \textsc{III}]$\lambda$1906/C \textsc{III}]$\lambda$1909} \citep[e.g.][]{Osterbrock06, Ke19_density}. \SII\ and \OII\ have lower critical densities and ionization energies than \mbox{C \textsc{III}]}, and therefore these tracers probe the gas conditions in different regions of the ionized nebulae \citep[e.g.][]{Acharyya19, Ke19_density}. In this work we focus on $n_e$ measurements made using the \SII\ and \OII\ doublet ratios.  
\end{sloppypar}

Emission line studies of strongly lensed galaxies at \mbox{$z\sim$~1.5~--~3} provided the first hints that \highz\ SFGs have significantly larger electron densities than local \HII\ regions \citep[e.g.][]{Hainline09, Bian10, Rigby11, Christensen12, Wuyts12_lens2, Wuyts12_lens, Bayliss14}. Subsequent spectroscopic surveys found that the typical $n_e$ in SFGs has decreased from \mbox{$n_e \sim$~200~--~300~\cmcubed} at $z\sim$~2~--~3 \citep[e.g.][]{Steidel14, Shimakawa15, Sanders16} to \mbox{$n_e \sim$~100~--~200~\cmcubed} at $z\sim$~1.5 \citep[e.g.][]{Liu08, Kaasinen17, Kashino17} and to \mbox{$n_e \sim$~30~\cmcubed} at $z\sim$~0 \citep[e.g.][]{HerreraCamus16, Kashino19}. However, the physical mechanism(s) responsible for driving this evolution are difficult to identify, and to date no quantitative models have been proposed to explain the density evolution.

\begin{sloppypar}
When interpreting $n_e$ measurements it is important to consider the geometry of the line-emitting material and the volume over which $n_e$ is measured. Consider an \HII\ region containing a collection of line-emitting structures with electron densities $n_{e, i}$, volumes $V_i$, and \SII\ luminosities $L_{{\rm [S II]},i}$. The \SIIb/\SIIr\ ratio probes the approximate line-flux-weighted average $n_e$ of these structures\footnote{This is true if the majority of the $n_{e,i}$ values fall in the regime where the relationship between $n_e$ and \SIIb/\SIIr\ is approximately linear; i.e. $n_e \simeq$~40~--~5000~\cmcubed\ \citep[e.g.][]{Osterbrock06, Ke19_density}.}; i.e. 
\begin{equation}
 n_e({\rm [S II]}) \simeq \sum_i{ \left( n_{e,i} \times L_{{\rm [S II]},i} \right)} ~/~ \sum_i{L_{{\rm [S II]},i}}
\end{equation}
The root-mean-square (rms) number of electrons per unit volume in the \HII\ region, also known as the rms electron density or $n_e$(rms), can be calculated from the \Ha\ luminosity and volume of the \HII\ region:
\begin{equation}
L({\rm H}\alpha {\rm , H II}) = \gamma_{H\alpha} V_{\rm H II}~n_e^2({\rm rms})
\label{eqn:lha_total}
\end{equation}
where $\gamma_{H\alpha}$ is the volume emissivity of \Ha\ \mbox{(3.56~$\times$~10$^{-25}~{\rm {\bf erg}~cm}^3~{\rm s}^{-1}$} for Case B recombination at 10$^4$~K). The total \Ha\ luminosity of this hypothetical \HII\ region can also be written as the sum of the \Ha\ luminosities of the individual line-emitting structures:
\begin{equation}
L({\rm H}\alpha {\rm , H II}) = \gamma_{H\alpha} \sum_i{ \left( V_i~n_{e, i}^2 \right)}
\label{eqn:lha_clumps}
\end{equation}
By combining Equations \ref{eqn:lha_total} and \ref{eqn:lha_clumps} we can derive an expression for the volume filling factor ($\mathit{ff}$) of these structures: 
\begin{equation}
\mathit{ff} \equiv \left( \sum_i~{V_i}\right) /~V_{\rm H II} = n_e^2({\rm rms}) \times \frac{\sum_i{V_i}}{\sum_i{ \left( n_{e,i}^2 \times V_i \right)}} 
\label{eqn:ff_def}
\end{equation}
Assuming that all of the line-emitting structures have roughly similar electron densities, and that the volume-weighted and light-weighted average densities are approximately equal, Equation \ref{eqn:ff_def} can be re-written as
\begin{equation}
\mathit{ff} \simeq \left[ n_e({\rm rms}) / n_e({\rm [S II]}) \right]^2
\label{eqn:ff}
\end{equation}
Observations of local \HII\ regions have found that $n_e$(\SII) and $n_e$(\OII) are much larger than $n_e$(rms), implying that the majority of the line emission originates from clumps with relatively low volume filling fractions of $\sim$~0.1~--~10\% \citep[e.g.][]{Osterbrock59, Kennicutt84, Elmegreen00, Hunt09, Cedres13}. It is therefore likely that the physical processes governing the ionized gas densities occur on spatial scales far below what can be resolved at \highz. However, global trends between $n_e$ and galaxy properties provide constraints on what types of physical processes are most likely to drive the evolution of the global, line-flux-weighted average $n_e$ in SFGs over cosmic time.
\end{sloppypar}

The electron density appears to be closely linked to the level of star formation in galaxies. \citet{Kaasinen17} found that there is no difference in the electron densities of galaxies at $z\sim$~0 and $z\sim$~1.5 when they are matched in SFR. The electron density has been found to correlate with specific SFR (sSFR) and SFR surface density (\sfrsd), at both low and high redshift \citep[e.g.][]{Shimakawa15, Bian16, Puglisi17, Jiang19, Kashino19}. There is also evidence for a spatial correlation between enhanced star formation activity and enhanced electron density in local galaxies \citep[e.g.][]{Westmoquette11, Westmoquette13, Mcleod15, HerreraCamus16, Kakkad18}.

Several scenarios have been proposed to explain the correlation between $n_e$ and the level of star formation. The initial $n_e$ is set by the density of the parent molecular cloud, which also determines \sfrsd\ through the Kennicutt-Schmidt relation. The radiation emitted by a star cluster dissociates and photo-ionizes the surrounding molecular gas to produce an \HII\ region with a local electron density of \mbox{$n_e \simeq 2 ~ n_{H_2}$} \citep[e.g.][]{Hunt09, Shimakawa15, Kashino19}. However, $n_e$ may change over time as a result of energy injection and/or \HII\ region expansion. The ambient density and pressure could significantly influence the dynamical evolution of \HII\ regions. \citet{Oey97,Oey98} proposed that \HII\ regions undergo energy conserving expansion powered by stellar winds and supernovae \citep[see also][]{Weaver77} until the internal pressure is on the order of the ambient pressure. \HII\ regions in denser environments may expand less, resulting in larger electron densities \citep[e.g.][]{Shirazi14, HerreraCamus16}. Another possibility is that \sfrsd, which sets the rate of energy injection by stellar winds and supernovae \citep[e.g.][]{ Ostriker11, Kim13}, may also govern the pressure and density in \HII\ regions \citep[e.g.][]{Groves08, Krumholz09, Kaasinen17, Jiang19}. Finally, it has been suggested that galaxies or regions with higher \sfrsd\ may have a larger fraction of young \HII\ regions which are still over-pressured with respect to their surroundings \citep[e.g.][]{HerreraCamus16, Jiang19}. It is important to note that while any of these scenarios could potentially explain a link between the level of star formation and the \textit{volume-averaged} electron density, the relationship between $n_e$(rms) and $n_e$(\SII) as a function of redshift has not yet been established observationally, largely due to the difficulty in determining the average luminosities and volumes of unresolved \HII\ regions.

Quantitative tests of these scenarios have also been hindered by the limited dynamic range of individual galaxy samples. Measurements of $n_e$(\SII) and $n_e$(\OII) in \highz\ galaxies have large associated uncertainties because the \SII\ and \OII\ emission lines are relatively weak, and the \OII\ doublet lines can be significantly blended in galaxies with large integrated line widths. In addition, the measurements could be biased by emission from ionized gas outflows, which are prevalent at \highz. The line-emitting gas in star formation driven outflows at $z\sim$~2 is $\sim$~5$\times$ denser than the line-emitting gas in the \HII\ regions of the galaxies driving the outflows \citep[e.g.][]{NMFS19}. In order to recover intrinsic correlations between galaxy properties and the electron densities in \HII\ regions, and to place stronger constraints on the physical driver(s) of the $n_e$ evolution, it is necessary to assemble a large sample of galaxies spanning a wide range in redshift and galaxy properties, while also minimizing the degree of contamination from line emission produced outside of \HII\ regions.

In this paper we use a sample of 611 galaxies with no evidence of AGN activity or broad line emission associated with outflows, drawn primarily from the \ktd\ \citep{Wisnioski15, Wisnioski19} and SAMI \citep{Bryant15, Scott18} integral field surveys, to investigate the physical processes driving the evolution of the typical electron density in SFGs from $z\sim$~2.6 to $z\sim$~0. The \ktd\ sample is distributed across three redshift bins at $z\sim$~0.9, $z\sim$~1.5 and $z\sim$~2.2, allowing us to examine the evolution of $n_e$ over \mbox{$\sim$~5 Gyr} of cosmic history with a single dataset. We apply the same sample selection, spectral extraction and stacking methodology to the SAMI sample to obtain a self-consistent measurement of $n_e$ at $z\lesssim$~0.1. The combined sample is centered on the star-forming MS at each redshift and spans more than three orders of magnitude in SFR. 

The paper is structured as follows. In Section \ref{sec:sample} we outline the properties of our galaxy samples and describe the methods used to stack spectra, measure the \SII\ doublet ratio and calculate the \HII\ region electron densities and pressures. We present our results on the redshift evolution of $n_e$(\SII), $n_e$(rms) and ionized gas filling factors in Section \ref{sec:redshift_evolution}, and explore how $n_e$(\SII) varies as a function of global galaxy properties in Section \ref{sec:ne_trends}. In Section \ref{sec:density_evolution_driver} we compare our density measurements to quantitative predictions for various potential drivers of the $n_e$ evolution, and evaluate the most likely causes of the elevated electron densities in SFGs at \highz. Our conclusions are summarized in Section \ref{sec:conclusions}.

Throughout this work we assume a flat $\Lambda$CDM cosmology with \mbox{H$_{0}$ = 70 \kms\ Mpc$^{-1}$} and \mbox{$\Omega_0$ = 0.3}. All galaxy properties have been derived assuming a \citet{Chabrier03} initial mass function.

\section{Data and Methodology}\label{sec:sample}
\subsection{\ktd+ Parent Sample}
The \highz\ SFG sample used in this paper is primarily drawn from the \ktd\ survey, a VLT/KMOS IFU survey focused on investigating the emission line properties of primarily mass-selected galaxies at \mbox{0.6~$< z <$~2.7} \citep{Wisnioski15, Wisnioski19}. The \ktd\ sample was drawn from the subset of 3D-HST galaxies with \mbox{log($M_*/M_\odot$)~$>$~9} and \mbox{$K_{\rm AB} \leq$~23~mag}, with the aim to achieve a homogeneous coverage of the star-forming population as a function of stellar mass and redshift. In this paper, we focus on the subset of 525 \ktd\ galaxies that were included in the \citet{NMFS19} study of outflows across the \highz\ galaxy population. These objects were selected to have \Ha\ emission detected at a signal-to-noise (S/N) per spectral channel $>$~3, and no strong telluric line contamination in the region around the \NII+\Ha\ complex. \citet{NMFS19} visually inspected the spectra of all galaxies to search for broad emission line components indicative of outflows, allowing us to isolate a sample of galaxies with no evidence of outflows for our analysis (see Section \ref{subsec:density_sample}).

We supplement our \ktd\ sample with galaxies from other \highz\ surveys that were also included in the \citet{NMFS19} analysis. 47 galaxies were drawn from the SINS/zC-SINF Survey \citep{NMFS09, NMFS18, Mancini11}, a VLT/SINFONI survey of 84 galaxies at \mbox{1.5 $< z <$ 2.5} selected on the basis of having secure spectroscopic redshifts and expected \Ha\ fluxes \mbox{$\geq$~5~$\times$~10$^{-17}$ erg s$^{-1}$ cm$^{-2}$}. Again, objects with low \Ha\ S/N or bad telluric contamination were excluded. Finally, we included six galaxies at \mbox{2 $< z <$~2.5} from the $K$ band selected sample of \citet{Kriek07, Kriek08} observed with VLT/SINFONI and Gemini/GNIRS, and the galaxy \mbox{EGS-13011166} at $z\sim$~1.5 observed with LBT/LUCI \citep{Genzel13, Genzel14}. Our combined \highz\ parent sample consists of 579 galaxies, of which $\sim$~90\% are drawn from \ktd, and therefore this sample is henceforth referred to as the \ktd+ parent sample.

\begin{figure*}
\centering
\includegraphics[scale=1.1, clip = True, trim = 0 210 0 0]{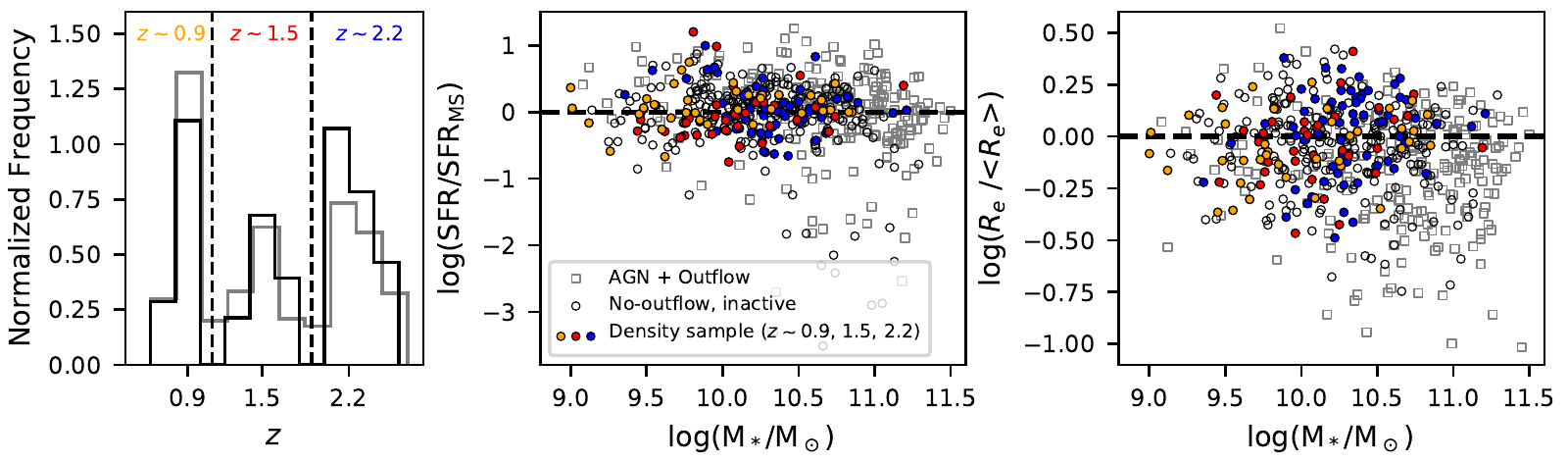}
\caption{Left: Normalized redshift distributions of the \ktd+ parent sample (grey) and our density sample (black). Center and right: Distribution of our density sample (solid markers) in (center) offset from the star-forming MS and (right) offset from the \citet{vanderWel14} galaxy mass-size relation, both as a function of stellar mass, compared to the distribution of the parent no-outflow inactive (open black circles) and AGN + outflow (open grey squares) samples. Orange, red and blue markers indicate galaxies in the $z\sim$~0.9, $z\sim$~1.5 and $z\sim$~2.2 redshift slices, respectively. \label{fig:sample_parameters}}
\end{figure*}

The grey histogram in the left-hand panel of Figure \ref{fig:sample_parameters} shows the redshift distribution of the \ktd+ parent sample. The galaxies are grouped in three distinct redshift slices, corresponding to the redshift ranges where \Ha\ falls into the KMOS $YJ$ ($z\sim$~0.9), $H$ ($z\sim$~1.5) and $K$ $(z\sim$~2.2) band filters.

Stellar masses were derived for all galaxies using population synthesis modeling of the rest-UV to optical/near-IR spectral energy distributions (SEDs), and SFRs were calculated from the rest-frame \mbox{UV + IR} luminosities using standard procedures, as described in \citet{Wuyts11b}. Galaxy stellar disk effective radii ($R_e$) were derived from two dimensional S\'ersic fits to \textit{HST $H$} band imaging \citep{vanderWel12, Lang14}. The properties of the \ktd\ and SINS/zC-SINF galaxies were taken directly from the survey papers which adopted the methods described above \citep{NMFS09, NMFS18, Mancini11, Tacchella15, Wisnioski19}.

\subsection{Extracting Integrated Spectra}\label{subsec:integrated_spectra}
\begin{sloppypar}
Integrated spectra for the \ktd\ and \mbox{SINS/zC-SINF} galaxies were extracted from the integral field datacubes as described in Section 2.5.1 of \citet{NMFS19}. Briefly, the datacubes were median subtracted to remove stellar continuum, 4$\sigma$-clipped blueward and redward of the strong emission lines to mask skyline residuals, and smoothed over the spatial dimensions using a Gaussian kernel with a typical full width at half maximum (FWHM) of 3 pixels for the KMOS cubes (0.6'') and 3~--~4 pixels for the SINFONI cubes (0.4~--~0.5'' for the seeing-limited datasets and 0.15~--~0.2'' for the adaptive optics assisted observations); comparable to the typical FWHM of the point spread function in all cases. A single Gaussian line profile was fit to the \Ha\ emission in each spaxel of the smoothed cubes to create velocity field maps, and the velocity field maps were used to shift the (unsmoothed) spectra of all spaxels within each galaxy to the same velocity centroid. The velocity shifting minimizes broadening of the integrated emission line profiles induced by the presence of large scale, gravitationally driven line-of-sight velocity gradients across rotating disks. Integrated spectra were extracted by summing the velocity shifted spectra of all spaxels within a galactocentric radius of \mbox{0.25~--~0.6''} (corresponding to a physical aperture radius of \mbox{2~--~5 kpc}, similar to the median $R_e$ of \mbox{3.4 kpc}), where the aperture size was adjusted based on the galaxy size to optimize the S/N of the extracted spectrum. 
\end{sloppypar}

\subsection{Selection of the \ktd+ Density Sample}\label{subsec:density_sample}
In this work, we focus on star-forming galaxies with no evidence of AGN activity or broad line emission indicative of outflows. \citet{NMFS19} created a single stacked spectrum of inactive galaxies with strong outflows spanning \mbox{0.6 $< z <$ 2.6}, and measured the \SII\ ratios and electron densities of the narrow ISM component and the broader outflow component individually. They found that the outflowing gas is significantly denser than the ISM material \citep[see also][]{Arribas14, Ho14, Perna17, Kakkad18, Fluetsch20}, suggesting that the ISM material may be shocked and compressed as it is swept up by the hot wind fluid. 

In principal, the typical $n_e$ in SFGs at each redshift could be measured by stacking the spectra of all galaxies (with and without outflows) and measuring the \SII\ ratio in the narrow line component. We construct such stacks for each redshift slice of the \ktd+ sample, but in the $z\sim$~0.9 and $z\sim$~1.5 stacks, the S/N of the broad outflow component  is not sufficient to permit a robust two-component decomposition of the emission line profiles (see Appendix \ref{appendix:twocomp_fitting}). If we fit only one kinematic component to each of the \SII\ lines, the measured electron density would be a line-flux-weighted average of the ISM density and the outflow density. Therefore, we remove galaxies with outflows prior to stacking. The potential impact of this choice on the measured electron densities is discussed at the end of this section.

AGN host galaxies are removed because 1) outflows are prevalent in AGN host galaxies \citep[e.g.][]{Genzel14, NMFS14, Harrison16, Leung19, NMFS19}, and 2) we calculate the electron density using \HII\ region photoionization models (discussed in Section \ref{subsec:density_calculations}), which cannot be applied to the spectra of AGN host galaxies because the AGN ionizing radiation field is significantly harder than an O star spectrum and will produce a very different ionization and temperature structure (see e.g. discussion in \citealt{Ke19_density, DaviesRic20}).

\citet{NMFS19} classified all galaxies in the \ktd+ sample as either AGN or inactive, and outflow or no-outflow. Galaxies were classified as AGN if their hard \mbox{X-ray} luminosity, radio luminosity, mid-IR colors, or \NIIHa\ ratio exceeded the threshold for pure star formation. Outflows were identified visually based on the presence of broad or asymmetric features in the integrated emission line profiles. The velocity shifting that was performed prior to spectral extraction increases the sharpness and S/N per spectral channel of the line emission from the galaxy disk (see e.g. Figure 1 of \citealt{Swinbank19}), and therefore maximizes the outflow detection fraction by pushing the detection limit to lower outflow velocities and mass outflow rates. The majority (356/579 or 61\%) of the galaxies were classified as inactive with no visually identifiable outflow component in the line emission (`no-outflow'). A further 87 galaxies (15\%) were classified as inactive with outflows, and the remaining 136 (23\%) galaxies were classified as AGN hosts (of which 94, or 16\% of the parent sample, have detected outflows). 

\begin{sloppypar}
Of the 356 inactive galaxies with no outflows, 320 have spectra covering the \SII\ doublet. The \SII\ emission lines are relatively weak (with a typical peak amplitude $\sim$5\% that of the \Ha\ line at \mbox{$z\sim$~1~--~2}), and small changes in the \SIIb/\SIIr\ ratio correspond to relatively large differences in the derived electron density, so it is very important to create a sample of spectra without significant sky contamination in the \SII\ doublet region. We visually inspected the spectra of all 320 no-outflow inactive galaxies and removed objects with elevated errors or bad systematics in the \SII\ region. This quality cut leaves us with a final sample of 140 galaxies (the `density sample'). 
\end{sloppypar}

The black histogram in the left-hand panel of Figure \ref{fig:sample_parameters} shows the redshift distribution of the density sample. Of our 140 galaxies, 39 galaxies fall in the $z\sim$~0.9 slice, 36 galaxies fall in the $z\sim$~1.5 slice, and 65 galaxies fall in the $z\sim$~2.2 slice. The density sample covers a wide redshift range and allows us to probe the $n_e$ evolution over $\sim$~5~Gyr in cosmic history with consistent data and analysis. 

The center and right-hand panels of Figure \ref{fig:sample_parameters} show how the galaxies are distributed in the $M_*-{\rm SFR}$ (center) and $M_*-R_e$ (right) planes. We have removed the average trends in SFR and $R_e$ as a function of stellar mass and redshift, adopting the \citet{Speagle14} parametrization of the star-forming MS (their Equation 28; chosen for consistency with the \citealt{Tacconi20} molecular gas depletion time scaling relation which is later used to estimate molecular gas masses) and the \citet{vanderWel14} mass-size relation for late type galaxies as a function of the Hubble parameter $H(z)$. The filled circles show the density sample (orange: $z\sim$~0.9, red: $z\sim$~1.5, blue: $z\sim$~2.2), the open circles show the no-outflow inactive galaxies that did not pass the visual inspection cut, and the open gray squares show galaxies with outflows and/or AGN activity. 

The density sample probes typical SFGs spanning \mbox{$\sim$~2 dex} in both $M_*$ and sSFR, and has a median stellar mass of \mbox{log($M_*/M_\odot$)~=~10.2}, with a slight trend towards higher stellar masses at higher redshift (the median stellar masses in the individual redshift bins are \mbox{log($M_*/M_\odot$)~=~9.9} at $z\sim$~0.9, \mbox{log($M_*/M_\odot$)~=~10.1} at $z\sim$~1.5, and \mbox{log($M_*/M_\odot$)~=~10.3} at $z\sim$~2.2). By nature of the selection criteria the density sample does not extend to the highest stellar masses or into the compact, quiescent and starburst galaxy regimes where AGN and outflows are most frequent \citep[see][]{NMFS19}. The removal of the highest stellar mass objects, which also have the highest SFRs, means that the density sample has a slightly lower median SFR than the parent sample at fixed $z$. The most actively star-forming galaxies are expected to have the highest $n_e$ \citep[e.g.][]{Shimakawa15, Kaasinen17, Jiang19, Kashino19}, and therefore there is a possibility that the electron densities measured from the density sample could under-estimate the true average $n_e$ in \HII\ regions at each redshift. However, we perform a test which suggests that the $n_e$ values measured from our density sample are likely to reflect the average gas conditions in \HII\ regions across the wider SFG population (see full description in Appendix \ref{appendix:twocomp_fitting}). 

\subsection{$z\sim$~0 Comparison Sample: SAMI Galaxy Survey}
We measure the zero-point of the $n_e$ evolution using a sample of galaxies from the SAMI Galaxy Survey \citep{Bryant15}, an integral field survey of $\sim$~3000 galaxies at \mbox{$z \lesssim$~0.1}. We choose an IFU sample rather than the much larger set of SDSS fiber spectra because the IFU data can be analysed using exactly the same methods applied to the \ktd+ data, allowing us to obtain a self-consistent measurement of $n_e$ at $z\sim$~0. We specifically choose the SAMI survey because 1) it is mass selected and 2) the spectral resolution ($R\sim$~4300) is similar to that of our \ktd+ data \mbox{($R\sim$~3500~--~4000)}. In comparison, the spectral resolution of the MaNGA survey is $R\sim$~2000 \citep{Bundy15}.

\begin{figure*}
\centering
\includegraphics[scale=1.1, clip = True, trim = 10 160 10 10]{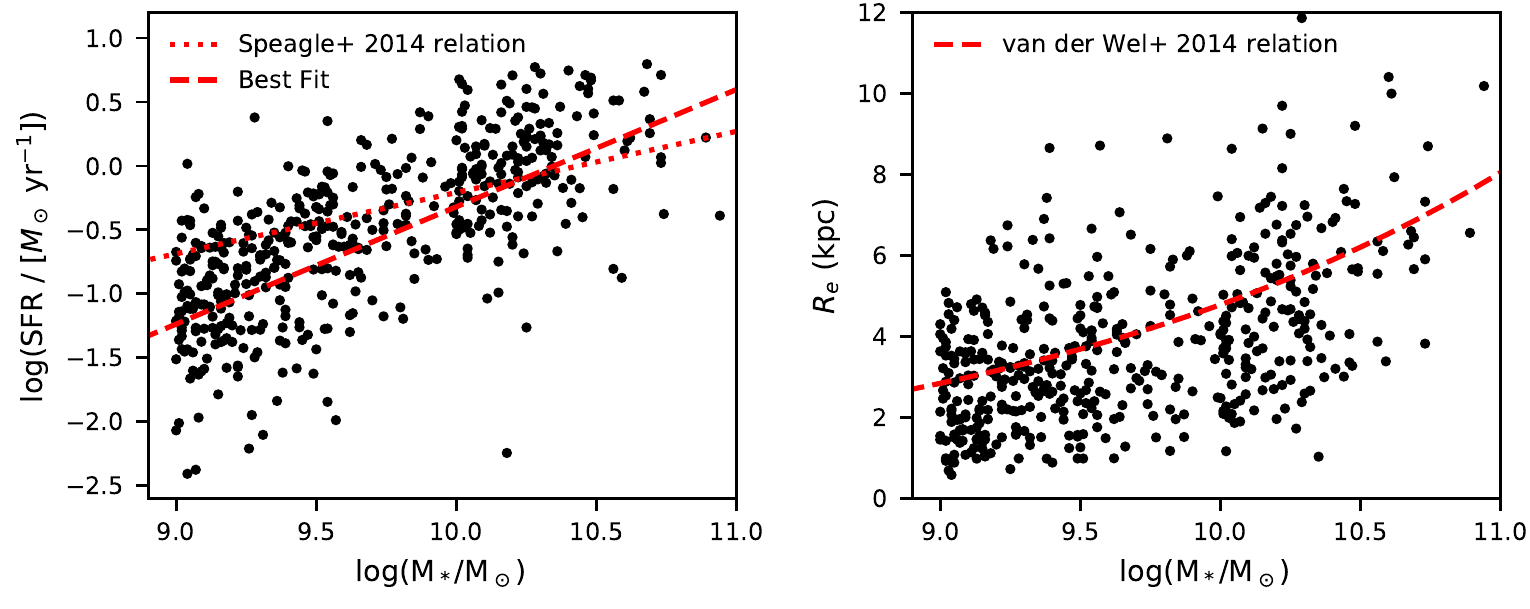}
\caption{Distribution of SAMI galaxies in the (left) $M_*-$~SFR and (right) $M_*-R_e$ planes. The red dashed line in the right-hand panel shows the $z\sim$~0 extrapolation of the \citet{vanderWel14} mass-size relation, which has been adjusted to the rest-frame central wavelength of the SDSS $r$ band filter ($\lambda \simeq$ 6020~\AA) using their Equation 1. \label{fig:sami_ms}}
\end{figure*}

\begin{sloppypar}
The most recent data release (DR2) includes blue and red data cubes (covering \mbox{3750~--~5750\AA} and \mbox{6300~--~7400\AA} observed, respectively) for 1559 galaxies, and velocity maps for 1526/1559 galaxies \citep{Scott18}. We start with 1197 galaxies that lie in the same stellar mass range as our \ktd+ targets (\mbox{$\log(M_*/M_\odot)$ = 9.0 -- 11.2}). Using the published emission line catalogues, we select 839 galaxies for which \Ha\ is detected at $\geq$~10$\sigma$ and \Hb, \NII$\lambda$6584 and \OIII$\lambda$5007 are all detected at $\geq$~3$\sigma$. We remove 280 galaxies with significant contributions from non-stellar sources (lying above the \citealt{Ka03} classification line on the \NIIHa\ vs. \OIIIHb\ diagnostic diagram). For each of the remaining 559 galaxies we velocity shift the blue and red datacubes, masking out spaxels for which no velocity measurement could be obtained, and then sum the velocity shifted cubes along both spatial dimensions to produce integrated spectra\footnote{The unmasked spaxels cover a median galactocentric radius of $\sim$~2$R_e$. This is larger than the typical radius covered by the \ktd+ spectra, but excluding spaxels outside 1$R_e$ does not have any significant impact on the electron densities measured from the SAMI spectra.}, as described in Section \ref{subsec:integrated_spectra}. Stellar continuum fitting and subtraction is performed by running the Penalized Pixel-Fitting (pPXF) method \citep{Cappellari04, Cappellari17} on the full (blue~+~red) spectrum for each galaxy, using the MILES library of stellar templates \citep{Vazdekis10}. The blue spectra are only used to constrain the continuum fitting and are not used in any further analysis. We visually inspect all integrated spectra and continuum fits, and reject galaxies with strong skyline residuals near any of the primary emission lines (\Ha, \NII\ and \SII), evidence for outflow emission (broad or asymmetric emission in multiple lines), or bad continuum fits. The final sample consists of 471 galaxies.
\end{sloppypar}

We calculate the global SFRs of the SAMI galaxies by summing the publicly available dust-corrected \Ha\ SFR maps (described in \citealt{Medling18}). The left-hand panel of Figure \ref{fig:sami_ms} shows how the SAMI galaxies are distributed in the $M_*-{\rm SFR}$ plane. The SAMI sample spans \mbox{$\sim$~2 dex} in $M_*$ and \mbox{$\sim$~3.5 dex} in sSFR, and has a median stellar mass of log($M_*/M_\odot$)~=~9.6; significantly lower than the median stellar mass of the \ktd+ sample (log($M_*/M_\odot$)~=~10.2) despite covering the same stellar mass range. The differences between the median stellar masses of the samples are accounted for when relevant to our analysis.

The red dotted line in Figure \ref{fig:sami_ms} shows the \citet{Speagle14} star-forming main sequence. The SAMI galaxies follow a slightly steeper relation indicated by the red dashed line, which is the best fit to the full sample of galaxies with \mbox{log(sSFR [yr$^{-1}$]) $>$ -11.2} (the approximate boundary between the star-forming and quiescent populations). The discrepancy in the main sequence slope is attributed to the fact that spaxels with significant contributions from non-stellar excitation sources are masked in the SAMI SFR maps, meaning that the calculated SFRs are lower limits \citep{Medling18}. Throughout the paper the main sequence offset of the SAMI galaxies is defined with respect to the best-fit (red dashed) line. 

The effective radii of the SAMI galaxies were derived from two dimensional S\'ersic fits to the GAMA $r$ band imaging \citep{Kelvin12}. The right-hand panel of Figure \ref{fig:sami_ms} shows where the SAMI galaxies lie in the \mbox{$M_*-R_e$ plane}, compared to the $z\sim$~0 extrapolation of the \citet{vanderWel14} mass-size relation (red dashed line) which has been adjusted to the rest-frame central wavelength of the SDSS $r$ band filter \mbox{($\lambda \simeq$ 6020~\AA)} using their Equation 1. The SAMI galaxies follow the expected increase in average size with increasing stellar mass but are $\sim$~10\% smaller than predicted by the \citet{vanderWel14} relation. 

\subsection{Stacking}\label{subsec:stacking}
We stack the integrated spectra of different sets of galaxies to produce high S/N composite spectra that can be used to make robust measurements of the \SII\ ratio, $n_e$, and thermal pressure. Before stacking, each spectrum is normalized to prevent the measured \SII\ ratios from being strongly biased towards galaxies with brighter line emission (i.e. galaxies at lower redshifts and/or with higher SFRs). The most accurate estimate of the average \SII\ ratio would be obtained by normalizing each spectrum to the peak amplitude of the \SIIr\ line, because this would, in the case of infinite S/N, yield the same result as measuring the \SII\ ratios of all galaxies individually and averaging the results. However, neither of the \SII\ lines is robustly detected in all of the galaxies. Instead we normalize to \Ha, which removes the majority of the variation in the \SIIr\ line amplitude because the SFR (which scales linearly with the \Ha\ luminosity) varies by 2~--~3 orders of magnitude within each redshift slice, whereas the \SIIHa\ ratios of galaxies with \HII-region-like spectra typically vary by only a factor of $\lesssim$~5 at fixed redshift \citep[e.g.][]{Ke06, Kashino17, Shapley19}.

The normalized galaxy spectra are averaged to obtain the stacked spectrum. When averaging, values lying more than 3$\sigma$ away from the median in each spectral channel are masked to ensure that the final stacks are not disproportionately affected by any possible remaining outliers.

\subsection{Electron Density and Thermal Pressure Calculations}\label{subsec:density_calculations}

\subsubsection{\SIIb/\SIIr\ Ratio and Model Grids}
\begin{sloppypar}
We measure the electron density and the thermal pressure from each stacked spectrum using the \SIIb/\SIIr\ ratio (also referred to as the `\SII\ ratio' and `\rSII'). \SIIb\ and \SIIr\ originate from excited states that have similar excitation energies but different collision strengths and radiative decay rates, meaning that the \SII\ ratio is strongly dependent on $n_e$ but only weakly dependent on temperature. In the low density limit, the timescale for collisional de-excitation is significantly longer than the timescale for radiative decay and the population ratio is determined by the ratio of the collision strengths, resulting in \rSII~$\sim$~1.45. In the high density limit, collisions govern transitions between the states and the electrons are distributed in a Boltzmann population ratio, resulting in \rSII~$\sim$~0.45. At densities similar to the critical density (where the probability of collisional de-excitation and radiative decay are approximately equal), \rSII\ varies almost linearly with $n_e$. The \SII\ ratio is most sensitive to densities in the range \mbox{$\sim$~40~--~5000~\cmcubed} \citep[e.g.][]{Osterbrock06, Ke19_density}, and is therefore a good probe of the electron density in the line-emitting material within \HII\ regions which typically ranges from tens to hundreds \cmcubed. 
\end{sloppypar}

We convert from \rSII\ to electron density and thermal pressure using the constant density and constant pressure model grids presented in \citet{Ke19_density}, respectively. The grids are outputs of plane-parallel \HII\ region models run with the \textsc{MAPPINGS} 5.1 photoionization code. The constant density models allow for a radially varying temperature and ionization structure within the nebula, and the constant pressure models additionally allow for radially varying density structure. Real \HII\ regions can have strong density gradients \citep[e.g.][]{Binette02, Phillips07} but are expected to have approximately constant pressure \citep[e.g.][]{Field65, Begelman90}, and therefore the pressure provides a more meaningful description of the conditions within \HII\ regions than the electron density. \footnote{We note that the electron densities derived from the outputs of the self-consistent \HII\ region photoionization models described here are generally in very good agreement with electron densities derived using model atom calculations that assume constant temperature and ionization structure, provided that the input atomic data are the same \citep{Ke19_density}.}

\begin{sloppypar}
Outputs of the constant density and constant pressure models are provided for \mbox{$\log(n_e/$\cmcubed) = 1.0 -- 5.0} and \mbox{$\log(P/k)$ = 4.0 -- 9.0}, respectively, with a sampling of 0.5 dex in both quantities. Throughout this paper, $P/k$ is in units of \mbox{$K$ \cmcubed}. For each value of $n_e$ and $\log(P/k)$, the grids include outputs of models run at five metallicities (\mbox{12 + log(O/H) = 7.63}, 8.23, 8.53, 8.93 and 9.23) and nine ionization parameters (\mbox{log $q$ = 6.5~--~8.5} in increments of 0.25 dex). The metallicity and ionization parameter determine the temperature structure of the nebula. The \SII\ ratio has a weak dependence on electron temperature because the collisional de-excitation rate scales with $T^{-1/2}$ (from the Maxwell-Boltzmann electron temperature distribution), and therefore the critical density scales with $T^{1/2}$ \cite[e.g.][]{Dopita03, Ke19_density}. 
\end{sloppypar}

\subsubsection{Measurements}\label{subsubsec:measurements}
We derive $n_e$(\SII) and \logpk(\SII) for each stacked spectrum by interpolating the model grids in $q$, $Z$ and \rSII. The \SII\ ratio is measured by fitting a single Gaussian to each of the \SII\ lines. We require both lines to have the same velocity centroid and velocity dispersion. 

We estimate the average metallicity of the galaxies in each stack using the \NII+\SII+\Ha\ calibration from \citet{Dopita16}. This diagnostic is relatively insensitive to variations in the density/pressure and ionization parameter, making it well suited for use with high redshift galaxies. \citet{Dopita16} calibrated the diagnostic using \textsc{MAPPINGS} 5.0 \HII\ region models run with the same abundance set as the \citet{Ke19_density} models, which is crucial because of the large systematic discrepancies between different metallicity calibrations in the literature. We simultaneously fit all the strong emission lines (\NII$\lambda$6548, \Ha, \NII$\lambda$6584, \SIIb, \SIIb) to measure the \NIIHa\ and \SIIHa\ ratios and obtain an estimate of the metallicity. The metallicity estimates for the \ktd+ stacks are listed in Table \ref{table:k3d_yjhk_measurements}.

Our \highz\ spectra do not cover the \OIII$\lambda$5007 and \OII$\lambda \lambda$3726,3729 emission lines which are required to make a direct measurement of the ionization parameter. We adopt typical ionization parameters of \mbox{log($q$)~=~7.8} for the \ktd+ galaxies based on measurements of star-forming galaxies at $z\sim$~1~--~2 from the COSMOS-\OII\ and MOSDEF surveys \citep{Sanders16, Kaasinen18}, and \mbox{log($q$) = 7.3} for the SAMI galaxies \citep{Poetrodjojo18}. However, varying the ionization parameter by a factor of three changes the derived pressures and densities by at most 0.1 dex (a factor of 1.2), and therefore the choice of ionization parameter has a minimal impact on our results.

We estimate the errors on the derived \rSII, $n_e$(\SII) and \logpk(\SII) values using a combination of bootstrapping and Monte Carlo sampling to account for both sample variance and measurement uncertainties. For a given stack of N galaxies, we randomly perturb the spectrum of each galaxy by its measurement errors, draw N perturbed spectra allowing for duplicates (bootstrapping), stack the drawn spectra, and measure \rSII, $n_e$(\SII) and \logpk(\SII). This process is repeated 600 times\footnote{This number was empirically verified to result in consistent error estimates between trials.}, and the 16th and 84th percentile values of the 600 measurements of \rSII, $n_e$(\SII) and \logpk(\SII) are taken as the lower and upper boundaries of the 1$\sigma$ confidence interval for each quantity. We note that due to the relatively high S/N of the input spectra, the error budget is dominated by sample variance in all cases.

\begin{figure*}
\centering
\includegraphics[scale=1.1, clip = True, trim = 0 185 0 0]{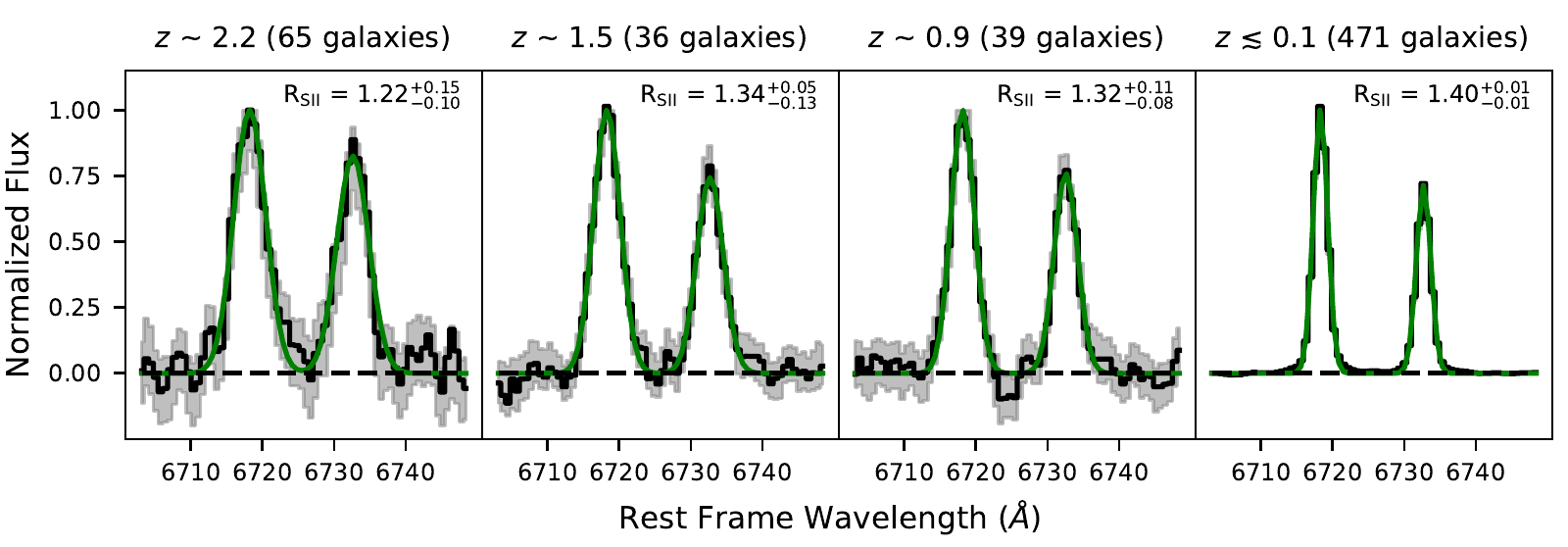}
\caption{Stacked \SII\ doublet profiles of galaxies with no evidence for outflows or AGN activity, in four redshift slices: \mbox{1.9 $\lesssim z \lesssim$~2.6}, \mbox{1.1 $\lesssim z \lesssim$ 1.9}, \mbox{0.6 $\lesssim z \lesssim$~1.1}, and $z\lesssim$~0.1. The grey shaded regions indicate the 1$\sigma$ spread of the 600 bootstrap stacks generated for each redshift slice (described in Section \ref{subsubsec:measurements}). The green curves indicate the best-fit Gaussian profiles. \label{fig:sfg_redshift_stacks}}
\end{figure*}

\section{Redshift Evolution of \HII\ Region Electron Densities}\label{sec:redshift_evolution}

\subsection{Typical \SII\ Electron Density at $z\sim$~0.9, $z\sim$~1.5 and $z\sim$~2.2 with \ktd+}\label{subsec:k3d_zslices}
We begin by using our `density sample' of inactive galaxies with no outflows to measure the average $n_e$(\SII) and \logpk(\SII) in each of the \ktd+ redshift slices. The stacked \SII\ doublet profiles and best Gaussian fits are shown in Figure \ref{fig:sfg_redshift_stacks}. The grey shaded regions indicate the 1$\sigma$ spread of the 600 bootstrap stacks generated for each redshift slice. 

\begin{figure*}
\centering
\includegraphics[scale=0.71, clip = True, trim = 0 0 170 40]{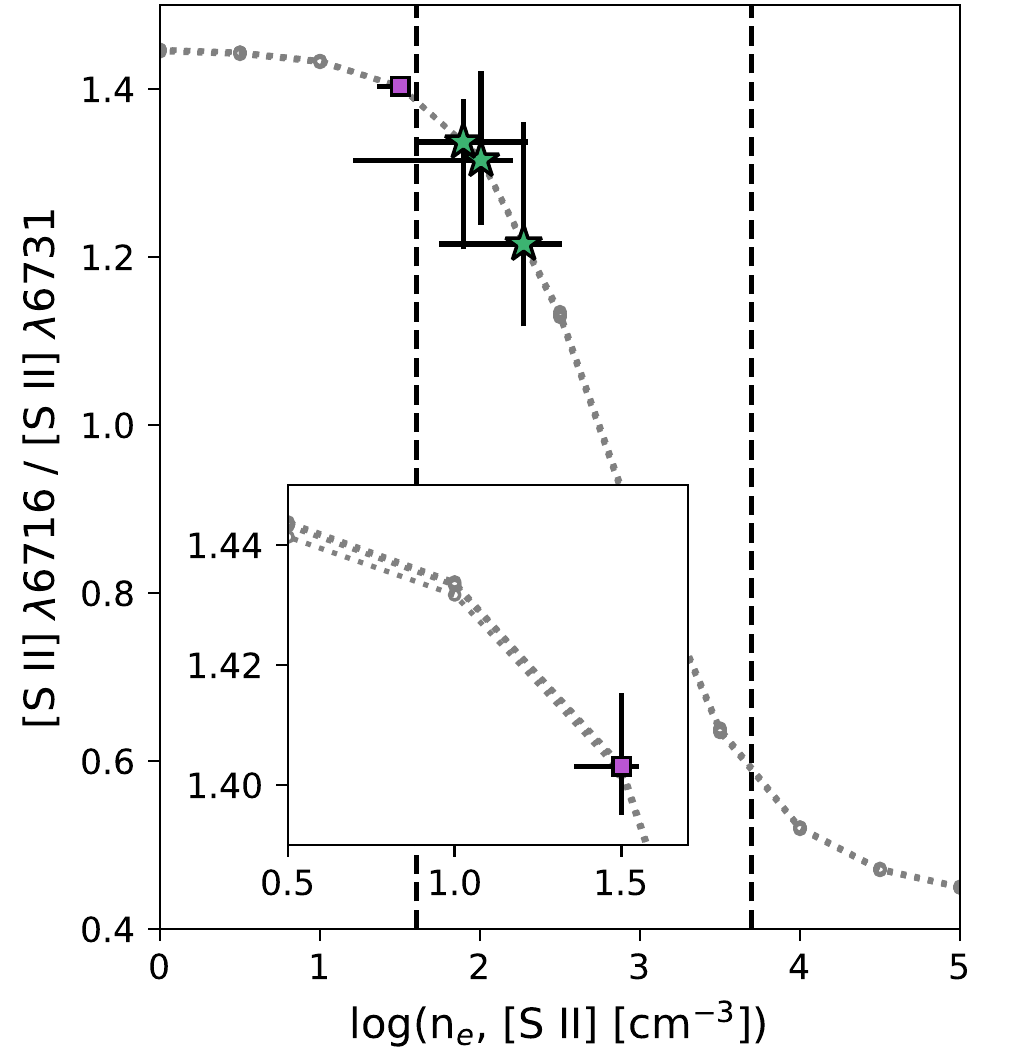}
\includegraphics[scale=1.3, clip = True, trim = 100 80 130 100]{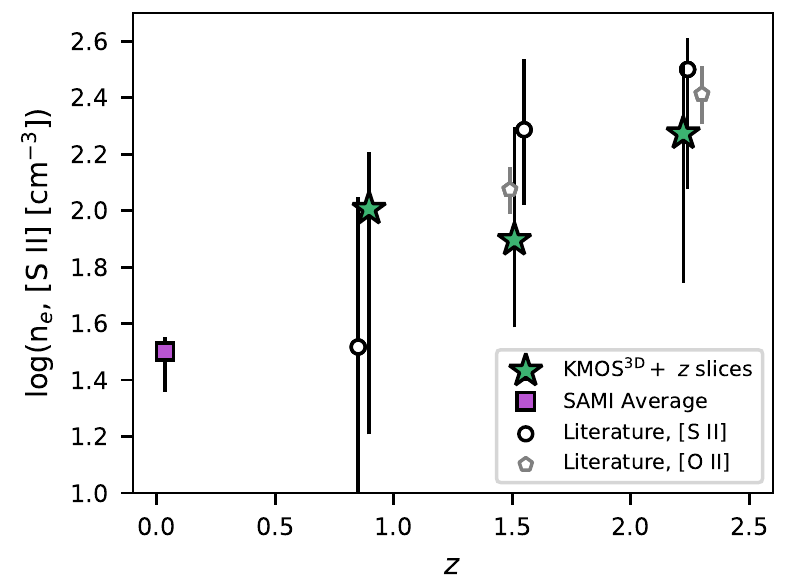}
\caption{Left: Illustration of the conversion between \rSII\ and $n_e$(\SII). For each redshift slice, we interpolate the \mbox{$q$~--~$Z$~--~$n_e$~--~\rSII} photoionization model output grid at the measured $Z$ and adopted $q$ to produce a set of ($n_e$, \rSII) pairs. The ($n_e$, \rSII) pairs are plotted as open grey circles, and the grey dashed lines are linear interpolations between the sampled electron densities. We generate and plot the circles and lines for each stack individually but the differences between the four sets of interpolated outputs are barely visible. Black dashed lines indicate the boundaries of the region where \rSII\ is most sensitive to $n_e$. The green stars and purple square show the measured \rSII\ and the derived $n_e$(\SII) for the \ktd+ and SAMI stacks, respectively. The inset in the bottom left is a zoom-in on the region around the SAMI measurement, showing that the measured \rSII\ is inconsistent with the theoretical maximum value. Right: The redshift evolution of $n_e$(\SII) based on the combination of the \ktd+, SAMI and \highz\ literature (\SII: black circles, \OII: grey pentagons) samples. \label{fig:density_evolution}} 
\end{figure*}

The left-hand panel of Figure \ref{fig:density_evolution} illustrates how the measured \SII\ ratios are converted to electron densities. For each redshift slice we interpolate the \mbox{$q$~--~$Z$~--~$n_e$~--~\rSII} photoionization model output grid at the measured $Z$ and adopted $q$ to produce a set of \mbox{($n_e$, \rSII)} pairs, plotted as grey circles. The grey dashed lines are linear interpolations between the sampled electron densities. We generate and plot the circles and lines for each stack individually but the differences between the sets of interpolated outputs are barely visible. The green stars and error bars show the \rSII\ measurements for the \ktd+ stacks and the corresponding $n_e$(\SII) values derived from the outputs of the constant density models. The \logpk(\SII) values are derived from the outputs of the constant pressure models using the same method. All of the measured and derived quantities are listed in Table \ref{table:k3d_yjhk_measurements}. 

\begin{sloppypar}
We find \mbox{$n_e$(\SII) = 101$^{+59}_{-85}$~\cmcubed} at $z\sim$~0.9, consistent with results from the KROSS survey \citep{Swinbank19}, and \mbox{$n_e$(\SII) = 79$^{+120}_{-40}$~\cmcubed} at $z\sim$~1.5, in agreement with measurements from the COSMOS-\OII\ \citep{Kaasinen17} and FMOS-COSMOS \citep{Kashino17} surveys. At $z\sim$~2.2 we measure \mbox{$n_e$(\SII)~=~187$^{+140}_{-132}$~\cmcubed}, similar to the values reported by the KBSS-MOSFIRE \citep{Steidel14} and MOSDEF \citep{Sanders16} surveys. 
\end{sloppypar}

The choice to remove galaxies with outflows from our sample was motivated by the observation of enhanced electron densities in outflowing material \citep{NMFS19}. However, the electron densities measured from our sample of no-outflow inactive galaxies match the electron densities measured from other galaxy samples that likely include star formation driven outflows. This suggests that the increased incidence of outflows at high redshift does not have a significant impact on the magnitude of the density evolution inferred from single component Gaussian fits to the \SII\ doublet lines. In Appendix \ref{appendix:outflow_densities} we confirm that including sources with star formation driven outflows (in proportion to their population fraction) has a minimal impact on the measured average densities. 

We also investigate the impact of AGN contamination on the measured densities. Uniform identification of AGN host galaxies at high redshift is challenging due to the varying availability and depth of multi-wavelength ancillary data between extragalactic deep fields. In Appendix \ref{appendix:agn_densities} we present tentative evidence to suggest that the measured densities could be up to a factor of $\sim$~2 larger when AGN host galaxies are included.

\subsection{Typical \SII\ Electron Density at $z\sim$~0}\label{subsec:ne_z0}
\begin{sloppypar}
We use the velocity shifted spectra of the sample of 471 SAMI galaxies to obtain a self-consistent measurement of the electron density at $z\sim$~0. The stacked \SII\ doublet profile and best Gaussian fit are shown in the right-most panel of Figure \ref{fig:sfg_redshift_stacks}. The purple square in the left-hand panel of Figure \ref{fig:density_evolution} indicates that the SAMI stack lies in the low $n_e$ regime of the \SII\ diagnostic where \rSII\ asymptotes towards the theoretical maximum value, causing the \mbox{$n_e$~--~\rSII} curve to become quite flat. However, the inset shows that due to the very high S/N of the stacked spectrum, the measured \rSII\ is inconsistent with the theoretical maximum value at the $\sim$~5$\sigma$ level, implying that we have a reliable measurement of $n_e$. 

The measured \rSII\ corresponds to an electron density of \mbox{$n_e$(\SII)~=~32$^{+4}_{-9}$~\cmcubed}. This value is in very good agreement with electron densities measured for resolved regions of local spiral galaxies using the \mbox{\NII122$\mu$m/\NII205$\mu$m} ratio which is a robust tracer of electron density down to \mbox{$n_e \sim$~10~\cmcubed} \citep{HerreraCamus16}, and with the typical $n_e$(\SII) derived from stacked SDSS fiber spectra of local galaxies \citep{Kashino19}.
\end{sloppypar}

\subsection{Redshift Evolution of the \SII\ Electron Density, and the Impact of Diffuse Ionized Gas}\label{subsec:ne_sii_evol}
We combine the SAMI and \ktd+ measurements to investigate how $n_e$ evolves as a function of redshift, as shown in the right-hand panel of Figure \ref{fig:density_evolution}. We also gather \SII\ and \OII\ ratio measurements from other surveys of \highz\ galaxies in the literature (KBSS-MOSFIRE; \citealt{Steidel14}, MOSDEF; \citealt{Sanders16}, KROSS; \citealt{Stott16}, COSMOS-\OII; \citealt{Kaasinen17}, and FMOS-KMOS; \citealt{Kashino17}). We require that the median SFR of each sample lies within 0.5 dex of the star-forming MS to ensure that the galaxies are representative of the underlying SFG population at the relevant redshifts. The majority of the literature measurements are based on slit spectra with the exception of the data from KROSS, a KMOS IFU survey of SFGs at \mbox{0.6 $< z <$ 1.0} \citep{Stott16}. A more complete description of the literature samples is given in Appendix \ref{sec:literature_samples}. We re-calculate the electron densities from the published line ratio measurements to avoid systematic biases in the conversion between line ratios and $n_e$ arising from differences in atomic data or assumed electron temperature (see e.g. discussions in \citealt{Sanders16} and \citealt{Ke19_density}). We do not calculate \logpk(\SII) for the literature samples because we do not have the line ratio measurements required to obtain self-consistent metallicity estimates.

The \SII\ and \OII\ lines originate from different regions of the nebula and will only give consistent densities if the electron temperature does not vary significantly between the \SII\ and \OII\ emitting regions. Sulfur exists as S$^+$ for photon energies in the range \mbox{10.4 -- 23.3 eV}\footnote{Ionization energies taken from the NIST Atomic Spectra Database (ver. 5.7.1); \href{https://physics.nist.gov/asd}{https://physics.nist.gov/asd}.}, and therefore \SII\ emission is expected to originate primarily from dense clumps and the partially ionized zone at the edge of the \HII\ region \citep[e.g.][]{Proxauf14}. On the other hand, O$^+$ exists for photon energies in the range \mbox{13.6 -- 35.1 eV} and therefore \OII\ is emitted over a much larger fraction of the \HII\ region \citep{Ke19_density}. However, \citet{Sanders16} showed that there is a good correspondence between the global \SII\ and \OII\ densities measured for star-forming galaxies at $z\sim$~2. In our plots, we distinguish between densities measured from the \SII\ ratio (circles with black outlines) and the \OII\ ratio (pentagons with grey outlines). 

Figure \ref{fig:density_evolution} clearly suggests that the typical electron densities inferred from the \SII\ and \OII\ doublet ratios have decreased by a factor of \mbox{$\sim$~6~--~10} over the last 10 Gyr, consistent with previous studies. However, to understand whether this reflects an evolution in the typical properties of ionized gas inside \HII\ regions, we must consider the origin of the line emission. It is well established that around 50\% of the \Ha\ emission from local galaxies originates from diffuse ionized gas (DIG) between \HII\ regions \citep[e.g.][]{Thilker02, Oey07, Poetrodjojo19, Chevance20}. The DIG is thought to be ionized by a combination of leaked ionizing photons from \HII\ regions, radiation from low mass evolved stars, and shock excitation \citep[e.g.][]{Martin97, RamirezBallinas14, Zhang17}. DIG dominated regions have larger \NIIHa\ and \SIIHa\ ratios than \HII\ regions \citep[e.g.][]{Rand98, Haffner99, Madsen06}, and it is therefore likely that a significant fraction of the \SII\ emission from the SAMI galaxies is associated with the DIG rather than \HII\ regions.

The line-emitting clumps in the DIG have a typical density of $n_e \sim$~0.05~\cmcubed\ \citep[e.g.][]{Reynolds91}, meaning that DIG contamination could potentially have a significant impact on the measured \SII\ ratios. Fortunately the \SII\ ratio saturates for densities below $n_e \sim$~40~\cmcubed\ (see the left-hand panel of Figure \ref{fig:density_evolution}), such that the \SII\ ratios measured for \HII\ regions at or below this density will be relatively unimpacted by DIG contamination. Recent surveys of resolved \HII\ regions in nearby spiral galaxies have found that the majority of \HII\ regions have \SII\ ratios in the low density limit \citep[e.g.][]{Cedres13, Berg15, Kreckel19}. In NGC~7793, the distributions of \SII\ ratios in \HII\ regions and the DIG are indistinguishable \citep{DellaBruna20}. These results suggest that the impact of DIG contamination on the derived $n_e$(\SII) at $z\sim$~0 may be relatively small. 

The situation is different at higher redshift where the measured electron densities are significantly above the low density limit of the \SII\ ratio. However, the fractional contribution of the DIG to the \Ha\ emission is anti-correlated with the \Ha\ surface brightness \citep[e.g.][]{Oey07}, and is predicted to decrease with increasing redshift until it becomes negligible at $z\sim$~2 (e.g. \citealt{Sanders17, Shapley19}, and see discussion in the following section). We therefore assume that the measured electron density evolution shown in Figure \ref{fig:density_evolution} is most likely to reflect a change in the intrinsic $n_e$(\SII) of \HII\ regions over cosmic time.

We note that even though DIG contamination is not expected to have a significant impact on the measured \rSII, the derived $n_e$(\SII) and \logpk(\SII) do not reflect the \textit{average} properties of gas in \HII\ regions. Galaxies commonly display negative radial gradients in \HII\ region electron density \citep[e.g.][]{Gutierrez10, Cedres13, HerreraCamus16} and metallicity \citep[e.g.][]{Zaritsky94, Moustakas10, Ho15}, the latter of which directly corresponds to positive electron temperature gradients because metal lines are the primary source of cooling in the 10$^4$~K ISM \citep[e.g.][]{Osterbrock06}. The derived $n_e$(\SII) and \logpk(\SII) represent the line-flux-weighted average properties of the gas within each aperture, and will therefore likely be biased towards the densest \HII\ regions in the central regions of the galaxies.

\subsection{Redshift Evolution of the Volume-Averaged Electron Density and Ionized Gas Filling Factor}\label{subsec:rms_density}

\subsubsection{Background}
The electron densities in \HII\ regions can be measured using two complementary methods. The measurements presented thus far have been based on \rSII, a density sensitive line ratio that probes the \textit{local} $n_e$ in the line-emitting material. The second approach is to use the \Ha\ luminosity, which is proportional to the volume emission measure, to calculate the rms number of electrons per unit volume, $n_e$(rms) (Equation \ref{eqn:lha_total}). The ratio of $n_e$(rms) to $n_e$(\SII) scales with the square root of the volume filling factor of the line-emitting material (Equation \ref{eqn:ff}).

There is some evidence to suggest that the rms electron densities (and by extension, the volume-averaged thermal pressures) of local \HII\ regions may be approximately proportional to the external ambient pressure \citep[e.g.][]{Elmegreen00, Gutierrez10}, hinting that the local environment may play an important role in regulating \HII\ region properties \citep[e.g.][]{Kennicutt84}. Measurements of $n_e$(rms) therefore represent a crucial link in our understanding of how global galaxy properties impact the local electron density of the line-emitting material.

The spatial resolution of our integral field observations is far below what is required to resolve individual \HII\ regions, and with our data we can only estimate the rms number of electrons per unit volume on galactic scales. This provides a lower limit on the rms number of electrons per unit volume within the \HII\ regions themselves, because \HII\ regions do not fill the entire volumes of star-forming disks. The rms electron density within the \HII\ regions is related to the measured $n_e$(rms) within $R_e$ by the inverse of the volume ratio: \mbox{$n_e$(rms, \HII) = $n_e$(rms, $R_e$) $\times~V_{\rm R_e} / V_{\rm H II}$}. The same scaling applies to the volume filling factors.

We use the SAMI and \ktd+ datasets to estimate $n_e$(rms) and the volume filling factor of the line-emitting gas within $R_e$ at $z\sim$~0, 0.9, 1.5 and 2.2. These calculations require a measurement of the \Ha\ luminosity within $R_e$ (described in Section \ref{subsubsec:dig_contribution}) and an estimate of the disk scale height (discussed in Section \ref{subsubsec:scale_heights}). 

\subsubsection{\Ha\ Luminosities}
\label{subsubsec:dig_contribution}
\begin{sloppypar}
The total \Ha\ luminosities within $R_e$ for the \ktd+\ galaxies are derived from the published integrated \Ha\ fluxes \citep{Kriek07, NMFS09, NMFS18, Wisnioski19} as follows. The \Ha\ fluxes are corrected for extinction using the continuum $A_V$ obtained from SED fitting and adopting the \citet{Wuyts13} prescription for extra attenuation towards nebular regions. The \Ha\ and $H$ band (observed frame) effective radii of SFGs at $z\sim$~1~--~2 are approximately equal \citep[e.g.][]{Nelson16, NMFS18, Wilman20}, and therefore we divide the integrated \Ha\ fluxes by two to obtain the fluxes within $R_e$. For the SAMI galaxies we directly use the published \Ha\ fluxes and dust correction factors within $R_e$ from the `recommend component' emission line flux catalogue \citep{Scott18}. The \Ha\ and $r$ band sizes of the SAMI galaxies are typically consistent to within $\sim$~0.1 dex \citep{Schaefer17}. 
\end{sloppypar}

The \Ha\ emission includes contributions from both \HII\ regions and DIG, as discussed in Section \ref{subsec:ne_sii_evol}. To isolate the \Ha\ emission from \HII\ regions, we assume that the fraction of \Ha\ emission associated with the DIG ($f_{{\rm H}\alpha {\rm , DIG}}$) follows the relationship calibrated by \citet{Sanders17}:
\begin{equation}
 f_{{\rm H}\alpha {\rm , DIG}} = -1.5 \times 10^{-14} \left(\frac{\Sigma_{H\alpha}}{{\rm erg~s}^{-1} {\rm~kpc}^{-2}}\right)^{1/3} + 0.748
 \label{eqn:fdig}
\end{equation}
This expression is the best fit to measurements of $\Sigma_{H\alpha}$ and $f_{{\rm H}\alpha {\rm , DIG}}$ for local galaxies. The power law index is fixed to 1/3, motivated by the assumption that there is a constant volume of gas available to be ionized, so that an increase in the total volume occupied by \HII\ regions (as a result of an increase in the SFR) directly corresponds to a decrease in the volume occupied by the DIG \citep{Oey07}. This assumption of density bounded ionization is likely to be unphysical because the implied escape fraction of ionizing photons from local starburst galaxies would be much larger than what is observed \citep{Oey07}. However, the functional form reproduces the general shape of the observed \mbox{$\Sigma_{H\alpha}-f_{{\rm H}\alpha {\rm , DIG}}$} trend.

Using Equation \ref{eqn:fdig} we estimate $f_{{\rm H}\alpha {\rm , DIG}} \simeq$~58\% at $z\sim$~0, $\sim$~33\% at $z\sim$~0.9, $\sim$~16\% at $z\sim$~1.5 and $\sim$~0\% at $z\sim$~2.2. The decrease in the estimated DIG contribution with increasing redshift is consistent with the \SIIHa\ ratios measured from our stacked spectra, which decrease from 0.38~$\pm$~0.01 at $z\sim$~0 to 0.19~$\pm$~0.01 at $z\sim$~2 \citep[see also][]{Shapley19}.

\subsubsection{Volume of the Star Forming Disk}
\label{subsubsec:scale_heights}
The volume of the star-forming disk within $R_e$ is computed assuming the disk is a cylinder with cross-sectional area $\pi R_e^2$ and height $2h_{\rm H II}$, where $h_{\rm H II}$ is the scale height of the star-forming disk. Ideally, $h_{\rm H II}$ would be directly measured from \Ha\ observations of edge-on disk galaxies. However, $z\sim$~0 disk galaxies show strong extraplanar \Ha\ emission associated with DIG \citep[e.g.][]{Rossa03, Miller03, Bizyaev17, Levy19}, meaning that the scale height of the star-forming disk cannot be measured from \Ha\ alone. 

A reasonable alternative is to take the typical scale height of the molecular gas disk out of which the \HII\ regions form, and correct this value upwards for the extra pressure support experienced by the ionized gas in the star-forming disk. The scale height and velocity dispersion of a thick and/or truncated gas disk are related by \mbox{$h \simeq R_d \times \sigma_0/v_{\rm rot}$} where $R_d$ is the disk scale length, $\sigma_0$ is the intrinsic velocity dispersion and $v_{\rm rot}$ is the rotational velocity \citep[e.g.][]{Genzel08}. Assuming that \HII\ regions and molecular clouds have similar radial distributions across galaxies, the kinematics and scale heights of the molecular and star-forming disks are related by
\begin{equation}
h_{\rm H II} = h_{\rm mol}~\times~(\sigma_0/v_{\rm rot})_{\rm H II}~/~(\sigma_0/v_{\rm rot})_{\rm mol}
\end{equation}
At fixed redshift, the typical velocity dispersion of ionized gas in SFGs is \mbox{$\sim$~10~--~15~\kms} larger than the average $\sigma_{\rm 0, mol}$ \citep{Uebler19}. The majority of this difference can be explained by the higher temperature of the ionized phase and the additional contribution of \HII\ region expansion to the measured velocity dispersion, which together are expected to contribute $\sim$~15~\kms\ in quadrature \citep[e.g.][]{Krumholz16}. Therefore, we assume that \mbox{$\sigma_{\rm 0, H II}$ = $(\sigma_{\rm 0, mol}^2 + 15^2)^{1/2}$}. 

Surveys of CO line emission in local spiral galaxies have found typical molecular gas velocity dispersions of \mbox{$\sim$~12~--~13~\kms} \citep{CalduPrimo13, Levy18}. We adopt \mbox{$\sigma_{\rm 0, mol}$~=~12.5~\kms}, from which we estimate \mbox{$\sigma_{\rm 0, H II}$~=~19.5~\kms}. The ionized gas is expected to have a slightly lower $v_{\rm rot}$ than the molecular gas because of the extra pressure support \citep[e.g.][]{Burkert10}, but the percentage difference is observed to be small \citep[e.g.][]{Levy18}, so we assume that \mbox{$v_{\rm rot, HII}/v_{\rm rot, mol} \simeq$ 1}. Molecular gas disks in the local universe have typical scale heights of 100~--~200~pc \citep[e.g.][]{Scoville93, Pety13, Kruijssen19}, so we adopt $h_{\rm mol}$~=~150~pc. Combining all these numbers, we estimate \mbox{$h_{\rm H II} \simeq$ 230~pc}.

At \highz\ the contribution of DIG to the \Ha\ emission is subdominant, but measurements of \Ha\ scale heights are very challenging due to surface brightness dimming. \citet{Elmegreen17} measured an average rest-UV continuum scale height of \mbox{0.63~$\pm$~0.24 kpc} for galaxies at $z\sim$~2, suggesting that \highz\ disks are significantly thicker than their low-$z$ counterparts. This is consistent with the elevated ionized gas velocity dispersions in \highz\ disks \citep[e.g.][]{Genzel06, Genzel08, Wisnioski15, Johnson18, Uebler19}. 

\begin{figure*}
\centering
\includegraphics[scale=1.1, clip = True, trim = 0 155 0 0]{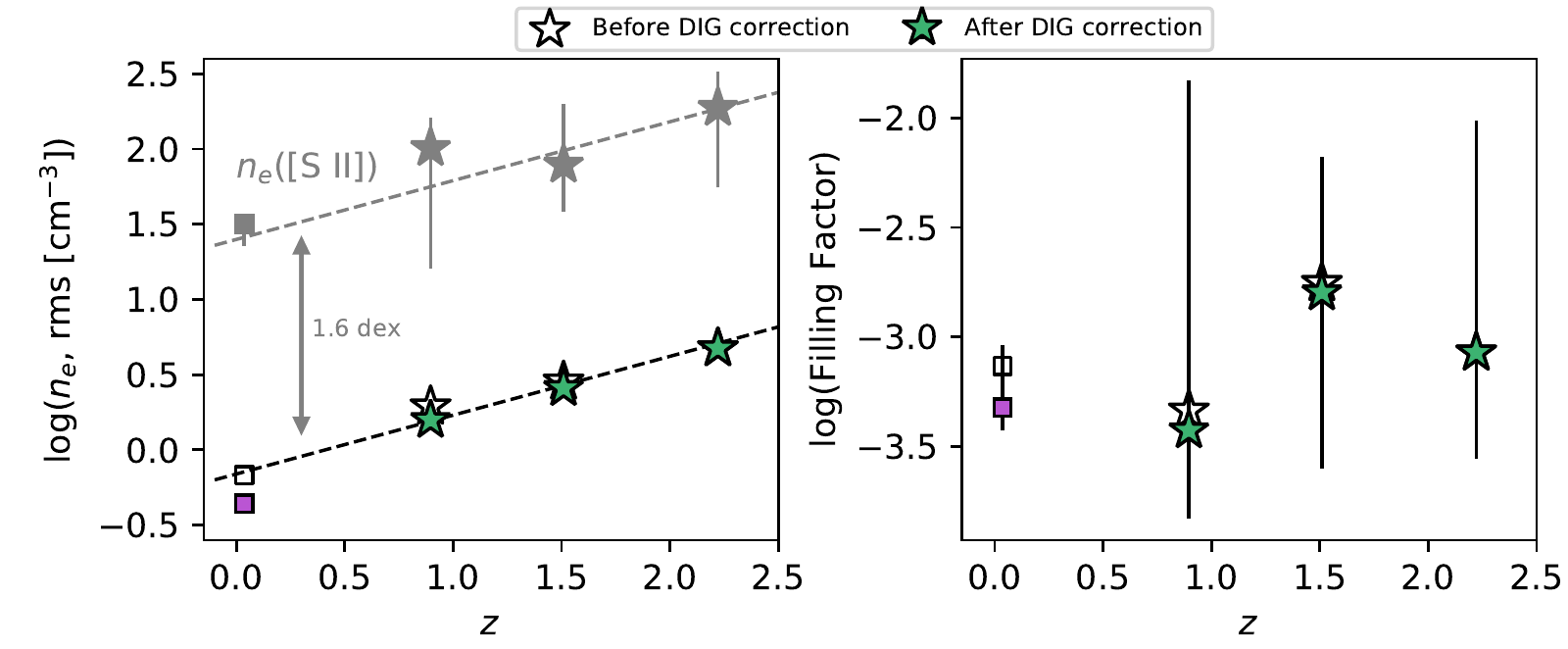}
\caption{Redshift evolution of the rms electron density (left) and the volume filling factor of the line-emitting gas (right). Grey markers in the left-hand panel show the $n_e$(\SII) measurements for comparison. The plotting symbols are the same as in \mbox{Figure \ref{fig:density_evolution}}. Open and colored solid markers indicate values before and after correcting for the contribution of DIG to the \Ha\ emission, respectively. Error bars are omitted from the open markers for clarity. The formal errors on the $n_e$(rms) measurements are too small to be seen. The two dashed lines in the left-hand panel were obtained by simultaneously fitting the redshift evolution of $n_e$(rms) and $n_e$(\SII), forcing both to have the same slope. The best-fit intercepts are offset by 1.6 dex. \label{fig:ne_rms_filling_factor}}
\end{figure*}

\begin{table*}
\begin{nscenter}
\caption{Thermal pressure and electron density calculated from the \SII\ doublet ratio, root-mean-square (volume-averaged) electron density, and volume filling factor of the line-emitting gas in each of the four redshift slices.}\label{table:filling_factor}
\begin{tabular}{lcccccc}
\hline
Redshift Bin & \logpk(\SII) & $n_e$(\SII, \cmcubed) & \multicolumn{2}{c}{$n_e$(rms, \cmcubed)} & \multicolumn{2}{c}{$\mathit{ff}_{\rm R_e} \, \times$10$^3$} \\ 
& & & Original & DIG corrected & Original & DIG Corrected \\ \hline \hline
$z\lesssim$~0.1 & 5.78$^{+0.01}_{-0.19}$ & 32$^{+4}_{-9}$ &0.7 $\pm$ 0.1 & 0.4 $\pm$ 0.1 & 0.7$^{+0.7}_{-0.2}$ & 0.5$^{+0.4}_{-0.1}$ \\
$z\sim$~0.9 & 6.29$^{+0.28}_{-0.73}$ & 101$^{+59}_{-85}$ &1.9 $\pm$ 0.2 & 1.6 $\pm$ 0.1 & 0.5$^{+17.8}_{-0.3}$ & 0.4$^{+14.4}_{-0.2}$ \\
$z\sim$~1.5 & 6.23$^{+0.43}_{-0.26}$ & 79$^{+120}_{-40}$ &2.8 $\pm$ 0.3 & 2.6 $\pm$ 0.3 & 1.8$^{+5.6}_{-1.5}$ & 1.6$^{+5.0}_{-1.3}$ \\
$z\sim$~2.2 & 6.62$^{+0.20}_{-0.53}$ & 187$^{+140}_{-132}$ &4.7 $\pm$ 0.4 & 4.7 $\pm$ 0.4 & 0.8$^{+8.9}_{-0.6}$ & 0.8$^{+8.9}_{-0.6}$ \\ \hline
\end{tabular}
\end{nscenter}
\tablecomments{$P/k$ is in units of ${\rm K~cm}^{-3}$. The \SII\ doublet ratio traces the local properties of the line-emitting gas. The rms electron density gives the average number of electrons per unit volume over the star-forming disk within $R_e$. We provide the values before and after correcting the \Ha\ luminosities for the contribution of diffuse ionized gas, as described in Section \ref{subsubsec:dig_contribution}.}
\end{table*}

We use measurements of $R_d$, $v_{\rm rot}$ and $\sigma_0$ to estimate the median $h_{\rm H II}$ at $z\sim$~0.9, 1.5 and 2.2. The \Ha\ flux profiles are assumed to be approximately exponential (motivated by studies of SFGs at similar redshifts; e.g. \citealt{Nelson13, Nelson16b, Wilman20}), which implies that \mbox{$R_d = R_e / 1.67$}. The $v_{\rm rot}$ and $\sigma_0$ values are measured by forward modeling the one dimensional velocity and velocity dispersion profiles extracted along the kinematic major axis of each galaxy, accounting for instrumental effects, beam smearing, and pressure support as described in \citet{Uebler19}. Their sample of galaxies with reliable kinematic measurements includes 18/39 of the galaxies in our $z\sim$~0.9 stack, 13/36 galaxies in our $z\sim$~1.5 stack, and 16/65 galaxies in our $z\sim$~2.2 stack. We estimate $h_{\rm H II}$ for each galaxy that is included in both our density sample and the \citet{Uebler19} kinematic sample and then calculate the median $h_{\rm H II}$ for each redshift slice, yielding approximate ionized gas scale heights of 280~pc, 460~pc, and 540~pc at $z\sim$~0.9, 1.5, and 2.2, respectively. The $h_{\rm H II}$ estimated for the $z\sim$~2.2 sample is consistent with the rest-UV continuum scale heights measured by \citet{Elmegreen17} for galaxies at the same redshift.

\subsubsection{Results}
\label{subsubsec:rms_ne_measurements}
The calculated rms electron densities and volume filling factors are shown in Figure \ref{fig:ne_rms_filling_factor} and listed in Table \ref{table:filling_factor}. We give the values before and after correcting for the DIG contribution to indicate the magnitude of the correction, which is relatively small because $n_e$(rms) scales with $(1-f_{{\rm H}\alpha {\rm , DIG}})^{1/2}$.\footnote{We note that the rms density becomes lower after correcting for the DIG contribution, even though the DIG is less dense than the ionized gas in the \HII\ regions, because there is no adjustment in the adopted line-emitting volume.} The quoted errors on $n_e$(rms) indicate the standard error on the mean based on the \Ha\ flux uncertainties, but in reality the error is dominated by the unknown systematic uncertainty on the line-emitting volume. 

Figure \ref{fig:ne_rms_filling_factor} indicates that $n_e$(rms) evolves at a very similar rate to $n_e$(\SII), increasing by a factor of $\sim$~6~--~10 from $z\sim$~0 to $z\sim$~2.2. Consequently, our estimates suggest that there is no significant evolution of the volume filling factor over the probed redshift range. These conclusions hold independent of whether or not the DIG correction is applied.

The line-emitting volume is calculated assuming that $h_{\rm H II}$ does not vary as a function of galactocentric radius. However, observations of constant ionized gas velocity dispersions across \highz\ disks \citep[e.g.][]{Genzel06, Genzel11, Genzel17, Cresci09} suggest that the scale height may grow exponentially with increasing galactocentric radius \citep[e.g.][]{Burkert10}. If we adopted a flared geometry the derived line-emitting volume would increase by a factor of 1.35. This would have a negligible impact on the derived rms electron densities (which scale with $h_{\rm H II}^{-1/2}$) and a minor impact on the derived filling factors (which scale with $h_{\rm H II}^{-1}$). For the same reason, the relatively large uncertainties on the ionized gas scale heights have a limited impact on our results. A factor of 2 change in any or multiple of the adopted scale height values would not change the basic conclusion that $n_e$(rms) evolves much more rapidly than the ionized gas volume filling factor.

The consistency between the rate of evolution of $n_e$(\SII) and $n_e$(rms) seen in the left-hand panel of Figure \ref{fig:ne_rms_filling_factor} suggests that the filling factor of the line-emitting material inside \HII\ regions may be approximately constant over cosmic time. This finding considerably reduces one major uncertainty in our understanding of the physical processes linking the evolution of $n_e$(\SII) to the evolution of galaxy properties.

The similarity between the redshift evolution of $n_e$(\SII) and $n_e$(rms) also provides further evidence to suggest that we are indeed observing a change in the density of the ionized material within \HII\ regions over cosmic time. The \SII-emitting gas in an \HII\ region with a radial $n_e$ gradient will have a different $n_e$ distribution depending on whether the nebula is ionization bounded or density bounded. Galaxies in the local group contain both ionization and density bounded \HII\ regions \citep[e.g.][]{Pellegrini12}, and the elevated \OIII/\OII\ and \OIII/\Hb\ ratios characteristic of \highz\ Ly$\alpha$ emitters could potentially be explained by density bounded nebulae \citep[e.g.][]{Nakajima14}. In a density bounded nebula the partially ionized zone is truncated, meaning that the observed \SII\ emission would originate from material at smaller radii which could have a higher average $n_e$ compared to the ionization bounded case. It is therefore hypothetically possible that some or all of the measured $n_e$(\SII) evolution could be driven by a decrease in the fraction of ionization bounded regions with increasing redshift, rather than by a change in the average gas conditions within \HII\ regions. However, the strong evolution of $n_e$(rms) suggests that changing gas conditions are the dominant source of the observed $n_e$(\SII) evolution.

\section{Trends between Electron Density and Galaxy Properties}\label{sec:ne_trends}
We begin our investigation into the physical origin of the density evolution by exploring how the electron density varies as a function of various galaxy properties, first within the \ktd+\ sample (Section \ref{subsec:k3d_correlations}) and then using the extended dataset (Section \ref{subsec:extended_trends}).

\begin{table*}
\begin{nscenter}
\caption{\SIIb/\SIIr\ measurements for stacks of \ktd+ galaxies in bins below and above the median in various galaxy properties.}\label{table:k3d_median_rsii}
\begin{tabular}{lcccccc}
\hline Property & Median Value & Below Median & Above Median & Difference & Significance ($\sigma$) \\
& & \rSII\ & \rSII\ & & \\ \hline \hline
\sfrsd\ ($M_\odot$~yr$^{-1}$ kpc$^{-2}$) & 0.3 & 1.38$^{+0.06}_{-0.10}$ & 1.16$^{+0.09}_{-0.09}$ &-0.22 & 1.6 \\
SFR ($M_\odot$~yr$^{-1}$) & 23 & 1.37$^{+0.09}_{-0.07}$ & 1.21$^{+0.09}_{-0.12}$ &-0.16 & 1.4 \\
log($\Sigma_{\rm baryon}$/($M_\odot$ kpc$^{-2}$)) & 8.7 & 1.39$^{+0.07}_{-0.13}$ & 1.17$^{+0.11}_{-0.06}$ &-0.22 & 1.3 \\
log(sSFR/yr$^{-1}$) & -8.9 & 1.39$^{+0.04}_{-0.12}$ & 1.17$^{+0.14}_{-0.08}$ &-0.21 & 1.2 \\
log($\Sigma_{H_2}$/($M_\odot$ kpc$^{-2}$)) & 8.4 & 1.36$^{+0.04}_{-0.13}$ & 1.19$^{+0.09}_{-0.08}$ &-0.17 & 1.1 \\
log($M_{H_2}/M_*$) & 0.06 & 1.36$^{+0.04}_{-0.11}$ & 1.19$^{+0.12}_{-0.09}$ &-0.17 & 1.0 \\
log(SFR/SFR$_{{\rm MS}(z)}$) & 0.0 & 1.34$^{+0.03}_{-0.16}$ & 1.21$^{+0.12}_{-0.05}$ &-0.13 & 0.7 \\
log($M_*/M_\odot$) & 10.2 & 1.33$^{+0.08}_{-0.09}$ & 1.24$^{+0.09}_{-0.10}$ &-0.09 & 0.7 \\
$R_e$ (kpc) & 3.4 & 1.28$^{+0.08}_{-0.08}$ & 1.27$^{+0.10}_{-0.09}$ &-0.01 & 0.1 \\ \hline
\end{tabular}
\end{nscenter}
\tablecomments{Each bin contains 70 galaxies. The rows are sorted by decreasing significance of the difference between the \SII\ ratios measured for the below and above median bins.}
\end{table*}

\begin{table*}
\begin{nscenter}
\caption{Electron densities and ISM pressures calculated from the \rSII\ measurements presented in Table \ref{table:k3d_median_rsii}.}\label{table:k3d_median_densities_pressures}
\begin{tabular}{lcccccccccccc}
\hline Property & \multicolumn{2}{c}{Below Median} & \multicolumn{2}{c}{Above Median} & \multicolumn{2}{c}{Difference} & \multicolumn{2}{c}{Significance ($\sigma$)} \\ 
& \logpk\ & $n_e$ (\cmcubed) & \logpk\ & $n_e$ (\cmcubed) & \logpk\ & $n_e$ (\cmcubed) & \logpk\ & $n_e$ (\cmcubed) \\ \hline \hline 
\sfrsd\ ($M_\odot$~yr$^{-1}$ kpc$^{-2}$) & 6.03$^{+0.46}_{-1.25}$ & 44$^{+86 }_{-41}$ & 6.73$^{+0.19}_{-0.20}$ & 257$^{+138 }_{-109}$ & 0.70 & 213 & 1.4 & 1.5 \\
SFR ($M_\odot$~yr$^{-1}$) & 6.09$^{+0.33}_{-2.09}$ & 52$^{+64 }_{-51}$ & 6.62$^{+0.25}_{-0.27}$ & 193$^{+170 }_{-82}$ & 0.54 & 141 & 1.3 & 1.4 \\
log($\Sigma_{\rm baryon}$/($M_\odot$ kpc$^{-2}$)) & 5.96$^{+0.59}_{-1.96}$ & 38$^{+108 }_{-37}$ & 6.72$^{+0.13}_{-0.28}$ & 252$^{+98 }_{-121}$ & 0.76 & 214 & 1.2 & 1.3 \\
log(sSFR/yr$^{-1}$) & 5.94$^{+0.57}_{-0.52}$ & 39$^{+99 }_{-27}$ & 6.73$^{+0.17}_{-0.41}$ & 246$^{+125 }_{-141}$ & 0.79 & 207 & 1.1 & 1.2 \\
log($\Sigma_{H_2}$/($M_\odot$ kpc$^{-2}$)) & 6.13$^{+0.48}_{-0.35}$ & 58$^{+115 }_{-29}$ & 6.65$^{+0.17}_{-0.26}$ & 214$^{+125 }_{-91}$ & 0.52 & 156 & 1.0 & 1.1 \\
log($M_{H_2}/M_*$) & 6.09$^{+0.46}_{-0.29}$ & 55$^{+99 }_{-22}$ & 6.68$^{+0.20}_{-0.37}$ & 216$^{+143 }_{-113}$ & 0.59 & 161 & 1.0 & 1.1 \\
log(SFR/SFR$_{{\rm MS}(z)}$) & 6.19$^{+0.52}_{-0.14}$ & 72$^{+159 }_{-23}$ & 6.63$^{+0.10}_{-0.42}$ & 192$^{+63 }_{-109}$ & 0.44 & 120 & 0.7 & 0.6 \\
log($M_*/M_\odot$) & 6.26$^{+0.34}_{-0.56}$ & 85$^{+81 }_{-62}$ & 6.55$^{+0.22}_{-0.31}$ & 160$^{+130 }_{-70}$ & 0.28 & 75 & 0.6 & 0.7 \\
$R_e$ (kpc) & 6.46$^{+0.20}_{-0.38}$ & 128$^{+81 }_{-73}$ & 6.51$^{+0.20}_{-0.43}$ & 139$^{+100 }_{-84}$ & 0.05 & 11 & 0.1 & 0.1 \\ \hline
\end{tabular}
\end{nscenter}
\end{table*}

\subsection{Trends in Electron Density Within \ktd+}\label{subsec:k3d_correlations}
We explore which galaxy properties are most closely linked to the density variation within the \ktd+ sample by dividing the galaxies into two bins (below and above the median) in various star formation, gas and structural properties: $M_*$, SFR, sSFR, \sfrsd, offset from the star-forming main sequence (SFR/SFR$_{{\rm MS}(z)}$), molecular gas fraction $\mu_{H_2}$ \mbox{(= $M_{H_2}/M_*$)}, molecular gas mass surface density \gmsd, baryonic surface density $\Sigma_{\rm baryon}$, and \reff. The molecular gas properties are estimated by assuming that the galaxies lie along the \citet{Tacconi20} scaling relation for the molecular gas depletion time $t_{\rm depl}$ as a function of $z$, offset from the star-forming MS and $M_*$. The derived $t_{\rm depl}$ is multiplied by SFR to obtain the molecular gas mass $M_{H_2}$. \sfrsd\ is defined as SFR/(2$\pi R_e^2$), and similar definitions apply to all other surface density quantities. The calculated surface densities represent the conditions in the central regions of the galaxies.

$\Sigma_{\rm baryon}$ is defined as the total surface density of the stellar, molecular gas, and atomic gas components. We adopt a constant atomic gas mass surface density of \mbox{$\Sigma_{HI}$~=~6.9~$M_\odot$~pc$^{-2}$} based on the tight observed relationship between the masses and diameters of \HI\ disks in the local universe \citep{Broeils97, Wang16}. The gas reservoirs in the star-forming disks of \highz\ SFGs are expected to be dominated by $H_2$ \citep[][and references therein]{Tacconi18, Tacconi20}, and therefore the assumption of a redshift-invariant $\Sigma_{HI}$ is unlikely to have any significant impact on the baryonic surface densities derived for the \ktd+ galaxies (see discussion in Appendix \ref{appendix:HI}).

We stack the spectra of the galaxies below and above the median in each property, and measure \rSII, $n_e$(\SII) and \logpk(\SII) for each stack. Table \ref{table:k3d_median_rsii} lists the median value of each galaxy property, the \SII\ ratios measured for the below median and above median stacks, the differences between the \SII\ ratios measured for the below and above median stacks, and the statistical significance of these differences. We note that the median values are calculated for each property individually and therefore the combination of these parameters does not necessarily represent a `typical' galaxy. Table \ref{table:k3d_median_densities_pressures} lists the $n_e$(\SII) and \logpk(\SII) values calculated from the \SII\ ratios listed in Table \ref{table:k3d_median_rsii}, as well as the differences and statistical significance of the differences between the values measured for the below and above median stacks. 

\begin{figure*}
\centering
\includegraphics[scale=1.1, clip = True, trim = 0 80 0 0]{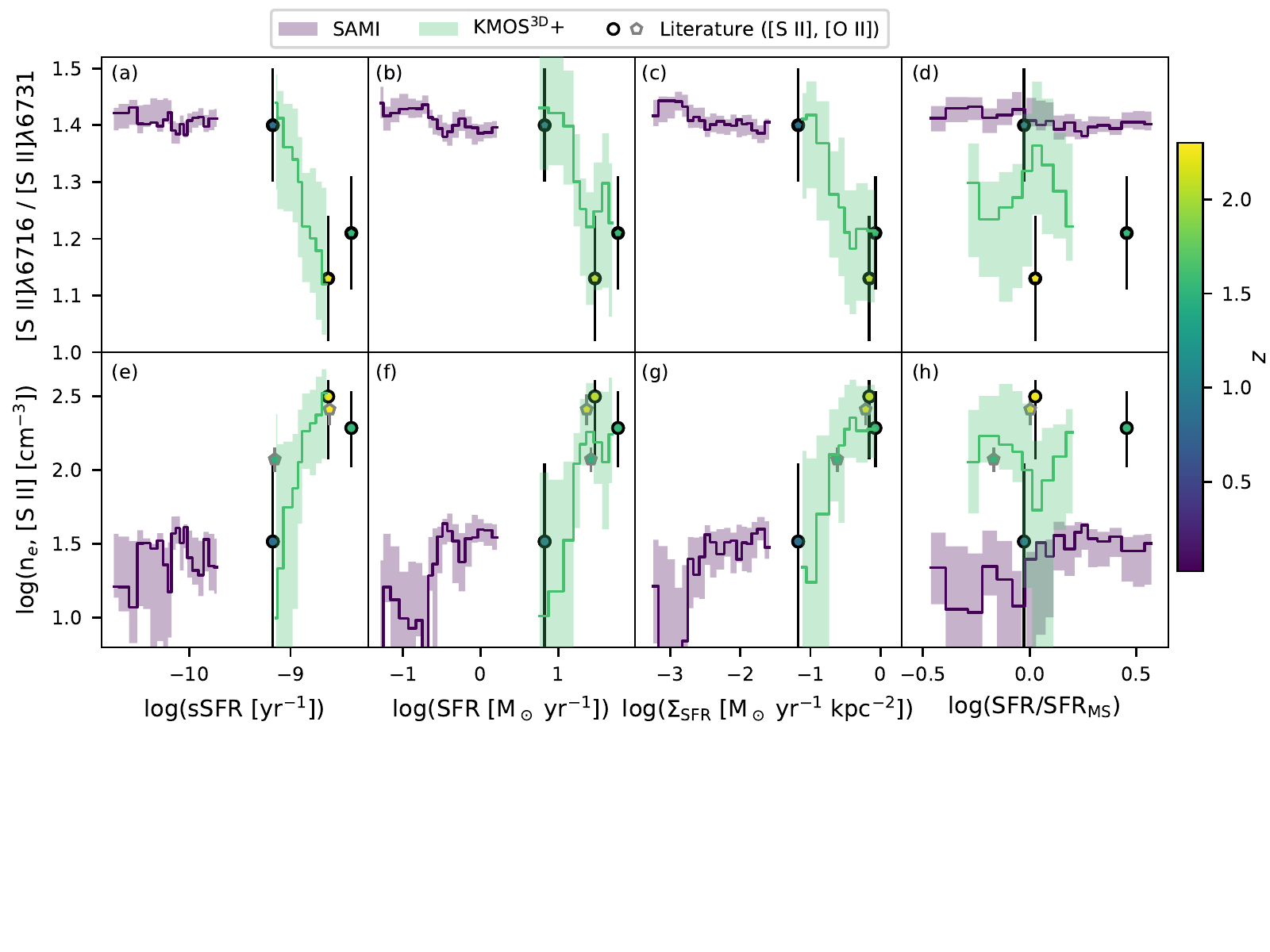} \\
\caption{\SII\ ratio (top) and $n_e$(\SII) (bottom) as a function of sSFR (a and e), SFR (b and f), \sfrsd\ (c and g) and SFR/SFR$_{{\rm MS}(z)}$ (d and h). The solid lines and shaded error regions represent the average properties of the SAMI (purple) and \ktd+ (green) galaxies, computed in sliding bins as described in Section \ref{subsec:extended_trends}. The individual data points are measurements for literature samples of \highz\ SFGs, introduced in Section \ref{subsec:ne_sii_evol}. In the bottom panel, $n_e$(\OII) measurements (points with grey outlines) have been included to illustrate the consistency between measurements made using different tracers. The color-coding indicates the median redshift of each galaxy sample.} \label{fig:rsii_sf} 
\end{figure*}

We find the most significant differences between the \SII\ ratios of galaxies below and above the median in \sfrsd, SFR, $\Sigma_{\rm baryon}$, and sSFR. The galaxies with comparably weak star formation and/or low $\Sigma_{\rm baryon}$ have electron densities and ISM pressures similar to those of local star-forming galaxies, whereas the galaxies with strong star formation and/or high $\Sigma_{\rm baryon}$ have densities and pressures which are comparable to or exceed the typical values for $z\sim$~2 SFGs (see Figure \ref{fig:sfg_redshift_stacks}). \rSII\ is also mildly anti-correlated with \gmsd\ and $\mu_{H_2}$. There is a weak trend towards lower \rSII\ at higher SFR/SFR$_{{\rm MS}(z)}$ and $M_*$, the latter of which is likely driven by the positive correlation between $M_*$ and SFR. Our results are consistent with previous findings that the electron density is positively correlated with the level of star formation in galaxies \citep[e.g.][]{Shimakawa15, Kaasinen17, Jiang19, Kashino19}. However, the trend with $\Sigma_{\rm baryon}$ suggests that the weight of the stars and ISM may also influence the density of the ionized gas in \HII\ regions.

Within the \ktd+ sample there is no evidence that the density is correlated with $R_e$, suggesting that the $n_e$(\SII) evolution is unlikely to be explained solely by the size evolution of galaxies. 

\subsection{Trends in Electron Density Across 0~$\lesssim z \lesssim$~2.6}\label{subsec:extended_trends}
We use the extended dataset to investigate the relationship between electron density and global galaxy properties over a much larger dynamic range. Figure \ref{fig:rsii_sf} shows how \rSII\ (top) and $n_e$(\SII) (bottom) vary as a function of sSFR, SFR, \sfrsd, and SFR/SFR$_{{\rm MS}(z)}$. The galaxy samples are color-coded by median redshift.

\begin{figure*}
\centering
\includegraphics[scale=1.1, clip = True, trim = 0 80 30 20]{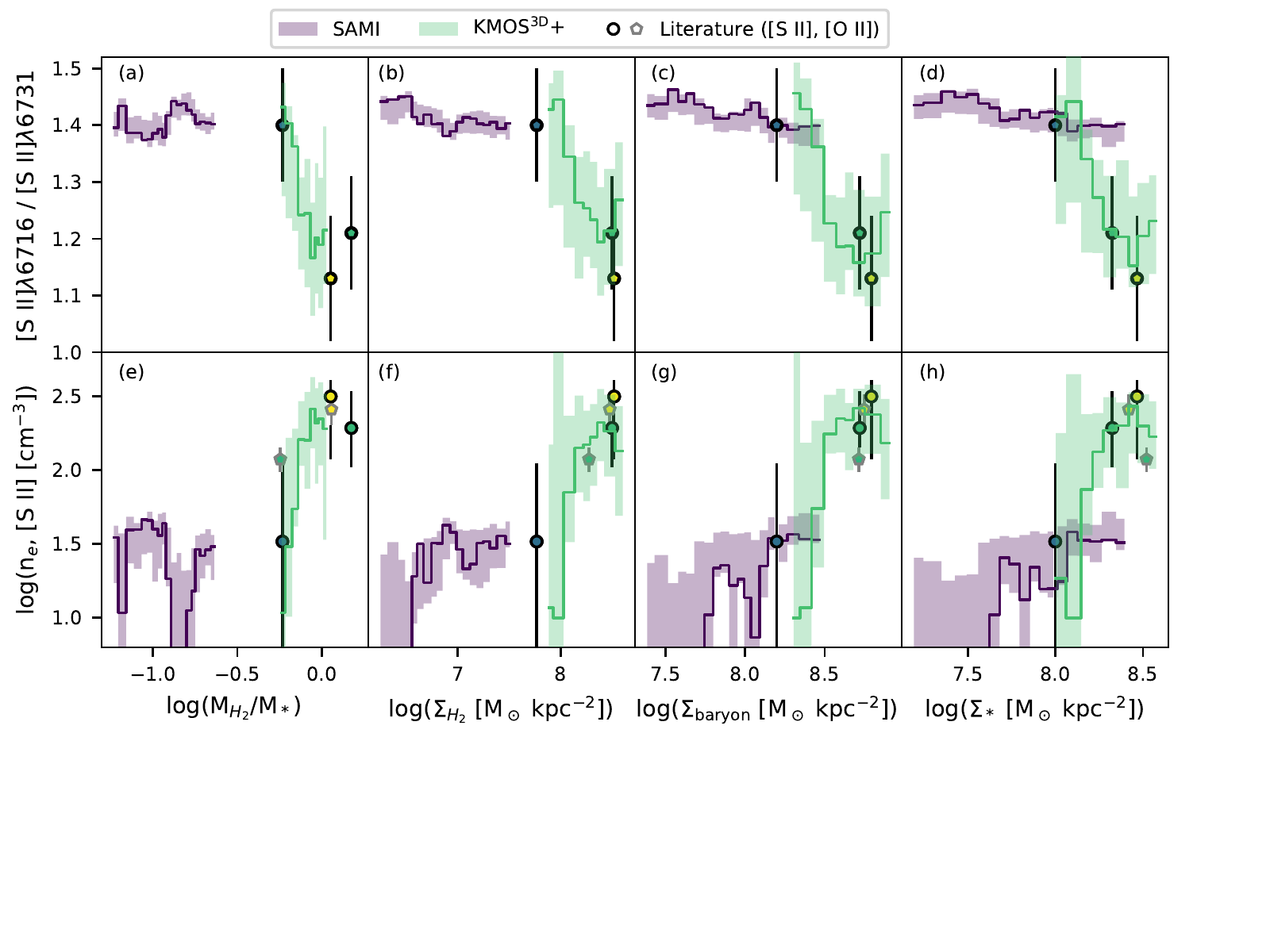} \\
\caption{\SII\ ratio (top) and $n_e$(\SII) (bottom) as a function of $\mu_{H_2}$ (a and e), \gmsd\ (b and f), $\Sigma_{\rm baryon}$ (c and g), and \smsd\ (d and h). The color-coding, symbols, and data representation are the same as in Figure \ref{fig:rsii_sf}.} \label{fig:rsii_stars_gas} 
\end{figure*}

We explore the trends within the SAMI and \ktd+ samples by measuring the \SII\ ratio in sliding bins. For the \ktd+ sample, we sort the galaxies by the quantity on the $x$-axis (e.g. sSFR), stack the first 50 galaxies, and calculate \rSII\ and $n_e$. The bin boundary is then moved across by 10 galaxies, and the stacking is repeated for galaxies number \mbox{10~--~60}, followed by galaxies number \mbox{20~--~70}, etc., resulting in a total of 9 bins. We perform measurements in sliding bins because it minimizes biases associated with the arbitrary choice of bin boundaries and gives a much clearer picture of the overall trends. However, the sliding bin measurements are highly correlated and are therefore not used in any quantitative analysis. 

The same procedure is applied to the SAMI galaxies, except that we stack in bins of 100 galaxies and move the bin boundary by 20 galaxies at a time, resulting in a total of 18 bins. The larger bin size is chosen to mitigate the effects of line ratio fluctuations in the low density limit. At the typical densities of the SAMI galaxies, \rSII\ changes very slowly as a function of $n_e$ (see left-hand panel of Figure \ref{fig:density_evolution}). Small line ratio fluctuations can lead to disproportionately large density fluctuations, which are partially smoothed out by the larger bins.

We find that the electron density is positively correlated with sSFR, SFR and \sfrsd, in good agreement with previous results \citep[e.g.][]{Shimakawa15, HerreraCamus16, Kashino19}. The trends among the \highz\ samples appear to be much steeper than the trends within the SAMI sample, but it is unclear whether this reflects an intrinsic difference in the relationship between $n_e$ and the level of star formation at different cosmic epochs, or whether it is an artefact of the flattening of the $n_e$~--~\rSII\ relationship at \mbox{$n_e \lesssim$ 40~\cmcubed}.

From Figure \ref{fig:rsii_sf}, it appears that the electron density is not intrinsically related to offset from the star-forming main sequence. The \ktd+ galaxies have systematically higher $n_e$(\SII) (lower \rSII) than the SAMI galaxies at fixed main sequence offset. This, combined with the overall roughly monotonic variations in \rSII\ and $n_e$ as a function of SFR, sSFR, and \sfrsd\ \citep[see also][]{Kaasinen17}, suggests that the redshift evolution of the electron density is likely to be linked to the evolving normalization of the star-forming MS.

Figure \ref{fig:rsii_stars_gas} shows \rSII\ (top) and $n_e$(\SII) (bottom) vary as a function of $\mu_{H_2}$, \gmsd, $\Sigma_{\rm baryon}$, and \smsd. $\mu_{H_2}$ is directly related to sSFR through the molecular gas depletion time and therefore the two quantities show very similar trends with $n_e$. The same is true for \gmsd\ and \sfrsd. Our data are consistent with a single positive correlation between $\Sigma_{\rm baryon}$ and $n_e$ across \mbox{0 $\lesssim z \lesssim$ 2.6}, but the \ktd+ galaxies are clearly offset to higher $n_e$(\SII) than the SAMI galaxies at fixed \smsd. This supports our earlier hypothesis that any correlation between $n_e$ and $M_*$ is primarily driven by the $M_*-$~SFR relation, and provides further evidence to suggest that the evolving gas content of galaxies -- which drives the evolution of the normalization of the star-forming MS -- may also be an important driver of the $n_e$ evolution.

\section{What Drives the Redshift Evolution of Galaxy Electron Densities?}\label{sec:density_evolution_driver}
We use our measurements to investigate possible physical driver(s) of the evolution of the electron density and thermal pressure across \mbox{0 $\lesssim z \lesssim$ 2.6}. We focus on four scenarios that are commonly discussed in the literature: that the electron density is governed by 1) the density of the parent molecular cloud (Section \ref{subsec:birth_cloud_density}), 2) the pressure injected by stellar feedback (Section \ref{subsec:feedback_pressure}), 3) the pressure of the ambient medium (Section \ref{subsec:ambient_pressure}), or 4) the dynamical evolution of the \HII\ region (Section \ref{subsec:stall_radius}). In this analysis we explicitly account for the average properties of the galaxies in each stack, meaning that the presented interpretation does not rely on the assumption that our samples are representative of the underlying SFG population at each redshift, or that the samples probe the same subset of the galaxy population at each redshift. 

\subsection{Scenario 1: \HII\ Region Density and Thermal Pressure Governed by Molecular Cloud Density}\label{subsec:birth_cloud_density}
Stars form in the centers of molecular clouds and radiate high energy photons that dissociate and ionize the surrounding ISM material to form \HII\ regions. Therefore, the initial electron densities of \HII\ regions are likely to be set by the molecular hydrogen number density. Since each $H_2$ molecule contributes two electrons, $n_e \simeq 2~n_{H_2}$.

The mass volume density of molecular hydrogen ($\rho_{H_2}$) within $R_e$ is derived by dividing \gmsd\ by the molecular gas scale height $h_{\rm mol}$, which is estimated using the procedures described in Section \ref{subsubsec:scale_heights}. We adopt $h_{\rm mol}$~=~150~pc at $z\sim$~0, and estimate the median $h_{\rm mol}$ at $z\sim$~0.9, 1.5 and 2.2 assuming \mbox{$h_{\rm mol} \simeq R_d \times \sigma_{0, \rm mol}/v_{\rm rot, mol}$}. The molecular gas kinematics are estimated from the measured ionized gas kinematics accounting for the expected difference in the pressure support experienced by the two gas phases \citep[e.g.][]{Burkert10}. These calculations yield typical molecular gas scale heights of $\sim$~180~pc, $\sim$~420~pc and $\sim$~490~pc at $z\sim$~0.9, 1.5 and 2.2, respectively. 

\begin{figure*}
\centering
\includegraphics[scale=1.1, clip = True, trim = 0 0 0 0]{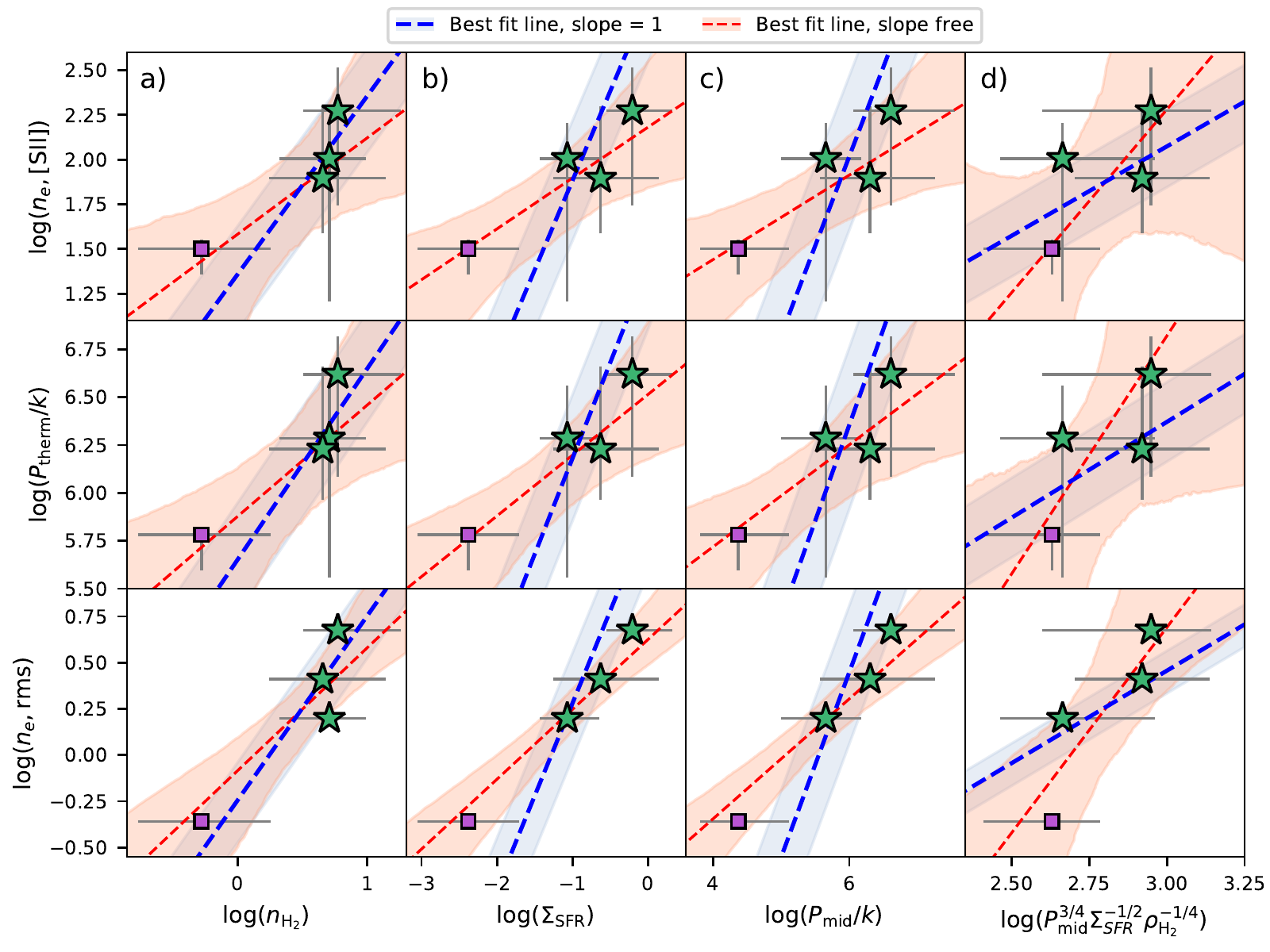} \\
\caption{Relationships between the thermal properties of the ionized gas -- probed by (top) $n_e$(\SII), (middle) \logpk(\SII), and (bottom) $n_e$(rms) -- and selected galaxy properties: a) $n_{H_2}$, b) \sfrsd, c) midplane pressure $P_{\rm mid}$, and \mbox{d) $\Sigma_{\rm SFR}^{-1/2}~\rho_{H_2}^{-1/4}~P_{\rm mid}^{3/4}$} which is proportional to the inverse of the predicted \HII\ region stall radius. The $n_e$(rms) measurements have been corrected for DIG contamination as described in Section \ref{subsubsec:dig_contribution}. Plotting symbols are the same as in Figure \ref{fig:density_evolution}. Error bars on the $x$ axis quantities indicate the 16th~--~84th percentile range in galaxy properties within each stack. The blue and red dashed lines are the best fits obtained when the slope is i) fixed to unity and ii) left free, respectively. The shaded regions show the 1$\sigma$ errors around the best fits, obtained by randomly perturbing each data point according to its errors and re-fitting 1000 times. When the blue and red lines match, the quantities on the $x$ and $y$ axes are consistent with having a 1:1 relationship (in log space) at all redshifts.} \label{fig:ne_scalings}
\end{figure*}

We divide $\rho_{H_2}$ by the molecular mass of $H_2$ to obtain $n_{H_2}$. Column a) of Figure \ref{fig:ne_scalings} compares the evolution of $n_{H_2}$ to the evolution of $n_e$(\SII) (top), \logpk(\SII) (middle), and $n_e$(rms) (bottom). The blue and red dashed lines are the best fits obtained when the slope is i) fixed to unity and ii) left free, respectively. The shaded regions indicate the 1$\sigma$ errors around the best fits, obtained by randomly perturbing each data point according to its errors and re-fitting 1000 times, and then computing the 16th~--~84th percentile range of these 1000 fits. A good match between the red and blue lines indicates that the quantities on the $x$ and $y$ axes are consistent with having a 1:1 relationship (in log space) at all redshifts. Any significant inconsistency between the blue and red lines in Column a) would suggest that the relationship between $n_e$ and $n_{H_2}$ changes over cosmic time, meaning that additional physical processes would need to be considered in order to explain the $n_e$ evolution.

\begin{sloppypar}
The blue and red lines in Column a) are very consistent with one another, indicating that there is an approximately linear relationship between $n_e$(rms) and $n_{H_2}$. The mean $n_e$(rms)/$n_{H_2}$ ratio across the four redshift slices is $\sim$~0.6. This relationship is averaged over the assumed volumes of the star-forming and molecular disks, and to obtain the average coefficient for individual star-forming regions one would need to multiply by the ratio of the volume filling factor of molecular clouds within 2$\pi R_e^2 h_{\rm mol}$ to the volume filling factor of \HII\ regions within 2$\pi R_e^2 h_{\rm H II}$. The discrepancy between the estimated coefficient of 0.6 and the predicted coefficient of 2 could very likely be accounted for by the systematic uncertainties introduced by the various assumptions made in our calculations. Therefore, we suggest that the elevated electron densities in \HII\ regions at \highz\ could plausibly be the direct result of larger molecular hydrogen densities in the parent molecular clouds.
\end{sloppypar} 

\subsection{Scenario 2: \HII\ Region Density and Thermal Pressure Governed by Stellar Feedback}\label{subsec:feedback_pressure}
\begin{sloppypar}
Although \HII\ regions are expected to form with \mbox{$n_e \simeq 2~n_{H_2}$}, the electron density may change over time as a result of energy injection and/or \HII\ region expansion. It has been suggested that the strong correlation between $n_e$ and the level of star formation in galaxies may arise because stellar feedback injects energy into \HII\ regions, increasing the internal pressure and electron density \citep[e.g.][]{Groves08, Krumholz09, Kaasinen17, Jiang19}.
\end{sloppypar}

\begin{sloppypar}
The turbulent pressure injected by stellar feedback can be parametrized as \mbox{$P_{\rm inj}$~=~\sfrsd\ ($p_*$/$m_*$)/4}, where $p_*$/$m_*$ is the amount of momentum injected into the ISM per solar mass of star formation, and the factor of 1/4 represents the fraction of the total momentum in the vertical component on one side of the disk \citep{Ostriker11, Kim13}. The value of $p_*$/$m_*$ scales with the number of supernovae per solar mass of star formation and is therefore strongly dependent on the initial mass function. ISM simulations have not yet reached a consensus on the amount of momentum injected per supernova explosion. It has been suggested that the momentum injection may be sensitive to small scale properties such as the spatial clustering of supernovae \citep[e.g.][]{Gentry19} and the interaction between the hot ejecta and the surrounding ISM \citep[e.g.][]{Kim15}, but differences in numerical methods lead to large discrepancies between different simulation results. \citet{Sun20} find a linear correlation between \sfrsd\ and the turbulent pressure of molecular clouds in local spiral galaxies, suggesting that $p_*/m_*$ is approximately constant. We assume that $p_*/m_*$ is constant and independent of redshift, meaning that \mbox{$P_{\rm inj} \propto \Sigma_{\rm SFR}$}. In other words, we can investigate the link between $n_e$ and pressure injection by stellar feedback without needing to assume a specific value for $p_*/m_*$.
\end{sloppypar}

Column b) of Figure \ref{fig:ne_scalings} indicates that the relationship between electron density and \sfrsd\ is likely to be significantly flatter than linear. This suggests that the increase in the rate of turbulent pressure injection towards higher redshifts does not directly lead to the observed increase in $n_e$. The sub-linearity of the relationship could potentially indicate that the fraction of injected pressure that is confined within \HII\ regions decreases towards higher redshifts. In order for the data to be consistent with a linear relationship between $n_e$ and the confined pressure, the fraction of pressure leaking out of \HII\ regions would have to increase by an order of magnitude from $z\sim$~0 to $z\sim$~2.2. The increased incidence of outflows at \highz\ \citep[e.g.][]{Steidel10, Newman12_global, NMFS19} could lead to a significant reduction in the pressure confinement efficiency, although our explicit removal of objects with detected galaxy scale outflows limits the possible magnitude of such an effect in our dataset.

Alternatively, the relationship between $n_e$ and the confined feedback pressure could be intrinsically sublinear, perhaps because a) the fraction of the injected turbulent pressure that cascades into the thermal pressure of the 10$^4$~K gas decreases steeply with increasing \sfrsd\ and/or $z$, b) $n_e$ is \textit{not} governed by the internal \HII\ region pressure, or c) $n_e$ \textit{is} governed by the internal pressure, but stellar feedback is not the primary source of internal pressure across some or all of the parameter space covered by our sample. The total pressure within an \HII\ region is the sum of many components including the turbulent and thermal pressure of the 10$^4$~K gas, the hot gas pressure associated with supernova ejecta and shocked stellar winds, and radiation pressure \citep[e.g.][]{Krumholz09, Murray10}. Quantitative predictions for the relationship between $n_e$ and $P_{\rm inj}$ from multi-phase simulations of \HII\ regions including outflows and all major pressure components \citep[e.g.][]{Rahner17, Rahner19}, as well as more complete observational censuses of the relative contributions of different pressure components within \HII\ regions \citep[e.g.][]{Lopez14, Mcleod19, Mcleod20}, would assist to determine which of these scenarios is most likely.

\subsection{Scenario 3: \HII\ Region Density and Thermal Pressure Governed by the Ambient Pressure}\label{subsec:ambient_pressure}
\HII\ regions form by dissociating and ionizing molecular gas, and are therefore initially over-pressured with respect to their surroundings. They expand towards lower pressures and densities until they reach equilibrium with the ambient medium. Analytic models suggest that in populations of \HII\ regions with average ages $\gtrsim$~1~Myr, the majority should be close to their equilibrium sizes and pressures \citep[e.g.][]{Oey97, Nath20}. 

The relationship between $n_e$(rms), $n_e$(\SII) and the pressure of the external ambient medium depends on the balance between the different internal pressure components. There is some observational evidence to suggest that in local \HII\ regions, $n_e$(rms) scales linearly with the disk midplane pressure. \citet{Elmegreen00} noted that the volume-averaged thermal pressures of the largest \HII\ regions in nearby massive spiral galaxies are comparable to the average disk midplane pressures (assuming \mbox{$T_e$~=~10$^4$~K}). \citet{Gutierrez10} measured $n_e$(rms) for individual \HII\ regions in a spiral galaxy and a dwarf irregular galaxy and found that in both galaxies, the electron density declines exponentially with a scale length similar to that of the \HI\ column density profile. The roughly linear relationship between $n_e$(rms) and the ambient pressure suggests that the thermal pressure of the 10$^4$~K gas must account for an approximately constant fraction of the total \HII\ region pressure.

If the majority of \HII\ regions at \mbox{0 $\leq z \leq$ 2.6} are in pressure equilibrium with their surroundings, and the balance between the different internal pressure components does not change significantly over time, then \mbox{$n_e$(rms) $\times~T_e$} should evolve at approximately the same rate as the midplane pressure. We do not expect the average $T_e$ to vary significantly between our four redshift slices because we measure similar gas-phase metallicities from all four stacks (the increase in median stellar mass towards higher redshift offsets the evolution of the mass-metallicity relation). Therefore, we estimate the average midplane pressure within $R_e$ at each redshift and test whether the midplane pressure evolves at a similar rate to $n_e$.

The pressure at the midplane of a disk in hydrostatic equilibrium is given by:
\begin{equation}
 P_{\rm mid} \simeq \frac{\pi G}{2} \Sigma_{HI+H_2} \left(\Sigma_{HI+H_2} + \frac{\sigma_g}{\sigma_*} \Sigma_* \right) \\
\end{equation}
where $\Sigma_*$ is the stellar mass surface density, $\Sigma_{HI+H_2}$ is the atomic + molecular gas mass surface density, $\sigma_g$ is the velocity dispersion of the neutral gas (which we assume to be given by $\sigma_{0, \rm mol}$), and $\sigma_*$ is the stellar velocity dispersion \citep{Elmegreen89}. The stellar velocity dispersion is estimated from \smsd\ assuming hydrostatic equilibrium, as outlined in Appendix \ref{appendix:sigma_stellar}. 

Column c) of Figure \ref{fig:ne_scalings} compares the evolution of $P_{\rm mid}$ to the evolution of the electron density and thermal pressure. Again, we find that the best-fit relationships have slopes significantly below unity, suggesting that the thermal pressure of the 10$^4$~K gas accounts for a decreasing fraction of the total \HII\ region pressure with increasing redshift. 

The implied change in the \HII\ region internal pressure balance can be understood by considering the sources of the different pressure components. The electron density increases by a factor of \mbox{$\sim$~6~--~10} from $z\sim$~0 to $z\sim$~2.2, which drives a similar change in the thermal pressure of the 10$^4$~K gas. The ratio of turbulent to thermal pressure in the 10$^4$~K gas is given by $\sigma_{\rm 0, ion}^2/c_s^2$. The sound speed $c_s$ scales with $T_e^{1/2}$ and is therefore not expected to vary significantly with redshift, but $\sigma_{\rm 0, ion}^2$ increases by a factor of $\sim$~4 from $z\sim$~0 to $z\sim$~2.2 \citep[e.g.][]{Genzel06, Wisnioski15, Johnson18, Uebler19}, driven by gravitational instabilities in marginally stable gas-rich disks \citep[e.g.][]{Krumholz18, Uebler19}. The radiation pressure and hot gas pressure both scale almost linearly with the average SFR \citep[e.g.][]{Oey97, Murray10, Ostriker11} which is approximately two orders of magnitude larger for a \mbox{$\log(M_*/M_\odot)$~=~10.3} galaxy at $z\sim$~2.2 than for a \mbox{$\log(M_*/M_\odot)$~=~9.6} galaxy at $z\sim$~0. Therefore, the ratio of the thermal pressure of the 10$^4$~K gas to the total \HII\ region pressure could easily change by at least an order of magnitude over the probed redshift range.

We conclude that $n_e$ could plausibly be set by the interplay between the external ambient pressure and the internal pressure balance, but not by the ambient pressure alone.

\subsection{Scenario 4: \HII\ Region Density and Thermal Pressure Governed by Dynamical Evolution}\label{subsec:stall_radius}
There is an inverse linear correlation between the diameter and $n_e$(rms) of local \HII\ regions \citep[e.g.][]{Kim01, Dopita06, Hunt09}, suggesting that the dynamical evolution of \HII\ regions may play an important role in regulating their electron densities. \citet{Oey97, Oey98} presented an analytic model for the size evolution of an \HII\ region assuming that the internal pressure is dominated by shocked stellar winds and supernova ejecta. They postulated that the \HII\ region will expand as an energy-driven bubble until the internal pressure is comparable to the ambient pressure, at which point the \HII\ region `stalls'. The stall radius scales as \mbox{$R_{\rm stall} \propto \Sigma_{\rm SFR}^{1/2}~\rho_{H_2}^{1/4}~P_{\rm mid}^{-3/4}$} (see derivation in Appendix \ref{appendix:rstall}). The final size of the \HII\ region is determined by the balance between the combined mechanical luminosity of the central star cluster and the internal gas pressure, which both drive the expansion of the \HII\ region, and the ambient pressure which resists the expansion.

\citet{Oey97} used their model to predict the slope of the \HII\ region size distribution, and showed that the majority of \HII\ regions are expected to be close to their maximum sizes. The predicted size distribution is consistent with the observed size distribution of \HI\ holes in local spiral galaxies \citep{Oey97, Bagetakos11}. If the majority of \HII\ regions at \mbox{0 $\leq z \leq$ 2.6} have sizes close to the stall radii predicted by \citet{Oey97}, and the size~--~$n_e$(rms) correlation observed in the local universe extends to higher redshifts, then we might expect to observe an inverse linear correlation between $n_e$(rms) and $R_{\rm stall}$, or equivalently, a (positive) linear correlation between $n_e$(rms) and $\Sigma_{\rm SFR}^{-1/2}~\rho_{H_2}^{-1/4}~P_{\rm mid}^{3/4}$.

We investigate whether or not such a correlation exists in Column d) of Figure \ref{fig:ne_scalings}. The relationship between $\Sigma_{\rm SFR}^{-1/2}~\rho_{H_2}^{-1/4}~P_{\rm mid}^{3/4}$ and $n_e$(rms) appears to be slightly super-linear, but the correlation between $\Sigma_{\rm SFR}^{-1/2}~\rho_{H_2}^{-1/4}~P_{\rm mid}^{3/4}$ and $n_e$(\SII) is approximately linear and could potentially be consistent with an inverse linear relationship between $n_e$ and $R_{\rm stall}$. This suggests that the elevated electron densities in \highz\ SFGs could plausibly be driven by a decrease in the ratio of the injected hot gas pressure (which drives \HII\ region expansion) to the midplane pressure (which resists the expansion).

However, we note that the adopted expression for the stall radius is based on a few simplifying assumptions which may not reflect the conditions in \HII\ regions at \highz. The most basic assumption is that each star cluster ionizes a spherical \HII\ region that does not overlap with regions ionized by adjacent clusters. Extreme star-forming regions in the local universe often show more complex geometries, with neighbouring star clusters ionizing overlapping areas within a single giant molecular cloud \citep[e.g.][]{NMFS01, Snijders07}. If these conglomerations of \HII\ regions reflect the typical conditions in rapidly star-forming galaxies at \highz, the stall radius may indicate the typical extent of the ionized region associated with an individual star cluster, rather than a characteristic spherical \HII\ region size.

The second key assumption is that the \HII\ region expansion is primarily driven by mechanical energy from stellar winds and supernovae. These processes produce highly pressurized hot gas which, if confined, can efficiently drive the expansion of an \HII\ region \citep[e.g.][]{Castor75}. However, observations of Milky Way \HII\ regions indicate that the pressure in the hot gas is comparable to the pressure in the 10$^4$~K gas \citep[e.g.][]{McKee84, Dorland87, HarperClark09}, suggesting that much of the hot gas may escape through holes in the expanding bubbles \citep[e.g.][]{Murray10}. If the hot gas pressure is significantly lower than predicted, the dynamics of \HII\ regions may be governed by other forms of pressure, such as the thermal pressure of the 10$^4$~K gas \citep[e.g.][]{Spitzer78, Dyson80} or radiation pressure. Photons trapped in the expanding shell of an \HII\ region will provide an extra source of acceleration, increasing the expansion velocity of the shell and the final stall radius. Radiation is predicted to govern the dynamics of \HII\ regions around luminous massive clusters \citep[e.g.][]{Krumholz09, Murray10} and may therefore be significant at \highz. However, thermal pressure and mechanical feedback are observed to be dominant in more typical local \HII\ regions \citep[e.g.][]{Lopez14, Mcleod19, Mcleod20}.

It is unclear how the relationship between $n_e$ and $R_{\rm stall}$ would change if a different expansion mechanism was considered. Stall radius expressions have been derived for thermal pressure driven expansion assuming that the expansion stalls when thermal pressure equilibrium is achieved \citep{Dyson80}, and for radiation pressure driven expansion assuming that the expansion stalls when the expansion velocity becomes comparable to the velocity dispersion in the parent cloud \citep{Krumholz09}. An investigation of how these stall conditions relate to the total pressure equilibrium condition for different ratios of thermal to total pressure is beyond the scope of this paper. Simulations of \HII\ region evolution including all major internal pressure components as well as varying fractions of turbulent pressure in the ambient medium (such as those presented in \citealt{Rahner17, Rahner19}) will help to better understand how the dynamical evolution and stall radii of \HII\ regions vary as a function of the luminosity and evolutionary stage of the cluster and the properties of the surrounding medium.

\subsection{Implications}\label{subsec:observational_comparison}
In this section we have compared the observed $n_e$ evolution to quantitative predictions for four potential drivers of the density evolution. We found that $n_e \simeq n_{H_2}$, suggesting that the elevated electron densities in \HII\ regions at high-$z$ could plausibly be the direct result of higher gas densities in the parent molecular clouds. We investigated whether the strong relationship between $n_e$ and the level of star formation in galaxies could arise because $n_e$ is governed by pressure injection from stellar feedback. Our data suggest that the increase in the amount of turbulent pressure injected by stellar winds and supernovae towards higher redshifts does not directly lead to the observed increase in $n_e$. Further constraints from observations and simulations are required to determine which additional parameters control the relationship between these two quantities. 

We explored whether the $n_e$ evolution could be driven by a change in the equilibrium internal pressure of \HII\ regions. We found that the ambient pressure evolves much faster than $n_e$, suggesting that $n_e$ could be governed by the interplay between the ambient pressure and the balance between different sources of pressure within \HII\ regions, but not by the ambient pressure alone. Finally, motivated by the existence of a strong inverse correlation between $n_e$(rms) and \HII\ region size in the local universe, we investigated whether the $n_e$ evolution could be linked to a change in the balance between the energy injection from stellar feedback (which drives \HII\ region expansion) and the ambient pressure (which resists the expansion). We found tentative evidence for an inverse linear correlation between $n_e$ and the \HII\ region stall radius $R_{\rm stall}$, but noted that the functional form of $R_{\rm stall}$ relies on some assumptions that may not hold at high-$z$. 

The evolution of $R_{\rm stall}$ is governed by the balance between the increase in \sfrsd\ and $\rho_{H_2}$ (which lead to larger $R_{\rm stall}$) and the increase in $P_{\rm mid}$ (which leads to smaller $R_{\rm stall}$). The fact that the predicted $R_{\rm stall}$ decreases with increasing redshift indicates that the $P_{\rm mid}$ term is dominant. Of the terms that contribute to $P_{\rm mid}$, the typical values of $\Sigma_{HI}$ and $\sigma_g/\sigma_*$ do not vary significantly across the redshift range probed by our samples, whereas the median \smsd\ in our $z\sim$~2.2 sample is $\sim$~0.7 dex higher than the median \smsd\ in our $z\sim$~0 sample, and the median \gmsd\ is 1.5 dex higher in our $z\sim$~2.2 sample than in our $z\sim$~0 sample. Therefore, the increase in $P_{\rm mid}$ is primarily driven by the increase in \gmsd. 

We conclude that, in the plausible scenario that the electron density is governed by either $n_{H_2}$ or 1/$R_{\rm stall}$, the increase in electron density from $z\sim$~0 to $z\sim$~2.6 would be primarily driven by the increase in the molecular gas fractions of galaxies. This is consistent with our earlier hypothesis that the $n_e$ evolution is linked to the evolving normalization of the star-forming MS (Section \ref{subsec:extended_trends}), which is also driven by the evolution of galaxy molecular gas reservoirs \citep[e.g.][]{Tacconi20}.

Finally, we note that the three rows of panels in Figure \ref{fig:ne_scalings} indicate that $n_e$(\SII), \logpk(\SII) and $n_e$(rms) evolve at very similar rates. The thermal pressure is expected to be approximately constant within any given \HII\ region and is therefore likely to be a more meaningful description of the gas conditions than $n_e$ which can show strong radial gradients (see e.g. \citealt{Ke19_density} and references therein). However, the conclusions presented in this section are not strongly dependent on which of the three quantities is considered.

\section{Conclusions}\label{sec:conclusions}
We have investigated the evolution of the typical electron density in SFGs from $z\sim$~2.6 to $z\sim$~0, using a sample of 140 galaxies at \mbox{0.6 $< z <$ 2.6} drawn primarily from the \ktd\ survey and a sample of 471 galaxies at $z\sim$~0 from the SAMI Galaxy Survey. The \ktd\ sample is distributed in three redshift bins ($z\sim$~0.9, 1.5, and 2.2) and allows us to analyse the density evolution over $\sim$~5~Gyr of cosmic history with a single dataset. We select galaxies that do not show evidence of AGN activity or broad line emission indicative of outflows in order to minimize contamination from line emission originating outside of star forming regions. We also examine the effects of diffuse ionized gas, which is expected to account for a decreasing fraction of the nebular line emission towards higher redshifts, and argue that this is unlikely to be the dominant driver of the observed redshift evolution of \rSII\ and of the derived $n_e$(\SII) and \logpk(\SII).

The galaxy spectra are stacked in bins of redshift and galaxy properties and the \SII\ doublet ratio is used to measure the local $n_e$ in the line-emitting gas. Based on these measurements, we find that:
\begin{itemize}
\item \begin{sloppypar} The electron density of the line-emitting gas in SFGs has decreased by a factor of $\sim$~6 over the last $\sim$~10~Gyr. We measure $n_e$(\SII)~=~187$^{+140 }_{-132}$~\cmcubed\ at $z\sim$~2.2, $n_e$(\SII)~=~79$^{+120 }_{-40}$~\cmcubed\ at $z\sim$~1.5, $n_e$(\SII)~=~101$^{+59 }_{-85}$~\cmcubed\ at $z\sim$~0.9, and $n_e$(\SII)~=~32$^{+4}_{-9}$~\cmcubed\ at $z\sim$~0, consistent with results from previous surveys of SFGs at similar redshifts. \end{sloppypar}
  
\item Combining the SAMI and \ktd+ datasets, we find that $n_e$(\SII) shows roughly monotonic correlations with sSFR, SFR and \sfrsd\ across \mbox{0 $\lesssim z \lesssim$ 2.6}. However, the \ktd+ galaxies have systematically higher $n_e$ than the SAMI galaxies at fixed offset from the star-forming MS, suggesting that the $n_e$ evolution is linked to the evolving main sequence normalization. There is also a roughly monotonic trend between $n_e$ and $\Sigma_{\rm baryon}$, but $n_e$ is correlated with $z$ at fixed $\Sigma_*$, suggesting that the gas reservoir plays an important role in regulating galaxy electron densities. 
\end{itemize}

We investigate how the $n_e$(\SII) measurements are impacted by contamination from non-stellar sources by comparing the \SII\ ratios measured from the spectra of galaxies with outflows and/or AGN activity to the \SII\ ratios measured for the primary sample. We measure higher $n_e$(\SII) for inactive galaxies with outflows than for no-outflow inactive galaxies, but only $\sim$~10~--~30\% of inactive SFGs at \mbox{0.6 $< z <$ 2.6} have outflows that are strong enough to be detectable in line emission, and this fraction is too low to have a significant impact on the measured average properties of the overall inactive SFG population. AGN host galaxies have lower \SII\ ratios than inactive SFGs and should be excluded to avoid over-estimating the average electron densities in star-forming regions.

We compare the local $n_e$(\SII) measurements to estimates of the rms number of electrons per unit volume across star-forming disks at each redshift. The rms electron density $n_e$(rms) is calculated from the \Ha\ luminosity (which is proportional to the volume emission measure) and the line-emitting volume (which we define as 2$\pi R_e^2 h$). The typical flattening ratios ($R_e/h$) of the high-$z$ disks are estimated from the measured $v_{\rm rot}/\sigma_0$ ratios. We find that $n_e$(rms) decreases by an order of magnitude from $z\sim$~2.2 to $z\sim$~0. The local and volume-averaged electron densities evolve at similar rates, suggesting that the volume filling factor of the line-emitting gas may be approximately constant across \mbox{0 $\lesssim z \lesssim$ 2.6}.
 
Finally, we use our measurements of $n_e$(\SII), \logpk(\SII), and $n_e$(rms) to explore different potential drivers of the $n_e$ evolution. We quantitatively test whether the electron density could plausibly be primarily governed by a) the density of the parent molecular cloud, \mbox{b) the} pressure injected by stellar feedback, c) the pressure of the ambient medium, or d) the dynamical evolution of the \HII\ region. We find that \mbox{$n_e$(rms) $\simeq n_{H_2}$}, suggesting that the elevated electron densities in \HII\ regions at \highz\ could perhaps be the direct result of higher gas densities in the parent molecular clouds. There is also tentative evidence to suggest that $n_e$ could be influenced by the balance between stellar feedback, which drives the expansion of \HII\ regions, and the ambient pressure, which resists their expansion.

Further studies are required to confirm the feasibility of these scenarios. Our molecular gas mass estimates are based on scaling relations and therefore the relationships between $n_e$, $n_{H_2}$, and $R_{\rm stall}$ should be verified using samples of galaxies with both optical spectroscopy and molecular gas measurements. The \HII\ region stall radii are estimated from analytic scalings which rely on many simplifying assumptions. Detailed comparisons between observed and predicted sizes for local \HII\ regions would help to establish whether such scalings can be meaningfully applied to predict the typical sizes of ionized regions in more distant galaxies.

Our conclusions fit with the growing picture that the evolution of the properties of SFGs from the peak epoch of star formation to the present day universe is primarily driven by a change in the rate of cold gas accretion onto galaxies. SFGs at $z\sim$~2 are thought to have elevated cold gas accretion rates, allowing them to maintain large molecular gas reservoirs which fuel rapid star formation, drive enhanced velocity dispersions, trigger the formation of massive clumps (see \citealt{Tacconi20} and F\"orster Schreiber et al. in press for reviews), and based on our work, may also be responsible for the elevated electron densities in \HII\ regions. \\

\acknowledgements
\begin{sloppypar}
We thank the referee for their constructive report which improved the clarity of this paper. RLD would like to thank Barbara Catinella and Brent Groves for informative discussions. EW and JTM acknowledge support by the Australian Research Council Centre of Excellence for All Sky Astrophysics in 3 Dimensions (ASTRO 3D), through project number CE170100013. MF acknowledges financial support from the European Research Council (ERC) under the European Union's Horizon 2020 research and innovation programme (grant agreement No 757535). DW acknowledges the support of the Deutsche Forschungsgemeinschaft via Projects WI 3871/1-1, and WI 3871/1-2. Based on observations collected at the European Organisation for Astronomical Research in the Southern Hemisphere under ESO Programme IDs 073.B-9018, \mbox{074.A-9011}, \mbox{075.A-0466}, \mbox{076.A-0527}, \mbox{077.A-0527}, 078.A-0600, \mbox{079.A-0341}, 080.A-0330, 080.A-0339, 080.A-0635, 081.A-0672, 081.B-0568, 082.A-0396, 183.A-0781, 087.A-0081, 088.A-0202, 088.A-0209, 090.A-0516, 091.A-0126, 092.A-0082, 092.A-0091, 093.A-0079, 093.A-0110, 093.A-0233, 094.A-0217, 094.A-0568, 095.A-0047, 096.A-0025, 097.A-0028, 098.A-0045, 099.A-0013, and 0100.A-0039. Also based on observations taken at the Large Binocular Telescope on Mt. Graham in Arizona. The LBT is an international collaboration among institutions in the United States, Italy and Germany. LBT Corporation partners are: The University of Arizona on behalf of the Arizona university system; Istituto Nazionale di Astrofisica, Italy; LBT Beteiligungsgesellschaft, Germany, representing the Max-Planck Society, the Astrophysical Institute Potsdam, and Heidelberg University; The Ohio State University, and The Research Corporation, on behalf of The University of Notre Dame, University of Minnesota and University of Virginia. This research made use of NASA's Astrophysics Data System. 
\end{sloppypar}

\software{\textsc{Astropy} \citep{Astropy13, Astropy18}, \textsc{emcee} \citep{ForemanMackey13}, \textsc{Matplotlib} \citep{Hunter07}, \textsc{Numpy} \citep{Oliphant06}}

\appendix

\section{Impact of Sample Selection on the Measured Electron Densities}
\subsection{Star Formation Rate Bias}\label{appendix:twocomp_fitting}
Our sample selection explicitly excludes AGN host galaxies and inactive galaxies with outflows because we are primarily interested in investigating what drives the evolution of the electron densities in \HII\ regions over cosmic time. However, a significant fraction of the excluded galaxies are located at high stellar masses and/or above the star-forming MS, and as a result, the density sample has a slightly lower median SFR than the parent sample at fixed $z$ (see Section \ref{subsec:density_sample}). The most actively star-forming galaxies are expected to have the highest $n_e$ \citep[e.g.][]{Shimakawa15, Kaasinen17, Jiang19, Kashino19}, and therefore the electron densities measured from the density sample could potentially under-estimate the true average $n_e$ in \HII\ regions of \mbox{$\log(M_*/M_\odot) \gtrsim$ 9 -- 9.5} galaxies at a given redshift. 

We investigate the impact of the small SFR bias on the derived electron densities by estimating the average $n_e$ of the narrow-line-emitting gas in different subsamples of galaxies with \SII-clean spectra. We produce stacks including no-outflow inactive galaxies, AGN hosts and/or inactive galaxies with outflows, and then fit the emission lines in the stacked spectra as superpositions of a narrow ISM component and a broader outflow component. The \SII\ ratio of the narrow component can be used to calculate the average $n_e$ in the disks of the stacked galaxies. However, a high S/N detection of the outflow component is required to obtain a meaningful two component decomposition of the emission line profiles. 

We perform the two component emission line fitting using \textsc{emcee}, a Markov Chain Monte Carlo (MCMC) Ensemble Sampler implemented in \textsc{Python} \citep{ForemanMackey13}. \textsc{emcee} returns the posterior probability distribution function (PDF) for each of the fit parameters and therefore allows us to evaluate whether or not the \SII\ ratio of the narrow component is well constrained by the data. Within each kinematic component, all emission lines are tied to the same velocity offset and dispersion. We fit all five emission lines (\NII$\lambda$6548, \Ha, \NII$\lambda$6583, \SII$\lambda$6716 and \SII$\lambda$6731) simultaneously to obtain the best possible constraints on the kinematics of the two components. We adopt flat priors on all fit parameters and impose a top-hat prior on the \SII\ ratio that has a value of 1 within the theoretically allowed range of \mbox{0.45 $\leq$ \rSII\ $\leq$ 1.45} and 0 outside of this range. The MCMC is run with 300 walkers, 300 burn-in steps and 1000 run steps. 

We obtain meaningful two component emission line decompositions for stacks of galaxies at $z\sim$~2.2. At lower redshifts, the outflow emission is not strong enough to break degeneracies between the fit parameters. For the $z\sim$~2.2 stacks, we use the posterior PDFs of the emission line narrow component amplitudes to derive the posterior PDF of $n_e$ following the method described in Section \ref{subsubsec:measurements}. From the stack of all inactive galaxies with \SII-clean spectra at $z\sim$~2.2 (96 galaxies) we measure a disk $n_e$ of \mbox{181$^{+123}_{-86}$~\cmcubed}, and from the stack additionally including AGN host galaxies (110 galaxies in total) we measure a disk $n_e$ of \mbox{207$^{+134}_{-100}$~\cmcubed}. These values are in very good agreement with the fiducial $n_e$ measured from the density sample \mbox{($n_e$~=~187$^{+140}_{-132}$~\cmcubed)}. \citet{NMFS19} performed similar two component line fitting on the stack of the 33 inactive outflow host galaxies with the highest S/N spectra across \mbox{0.6 $< z <$ 2.6} (from the \ktd+ sample). They reported a narrow \SII\ ratio of 1.33~$\pm$~0.09, almost identical to the ratio of 1.34~$\pm$~0.03 obtained from the single component fit to the stack of all no-outflow inactive galaxies in the same redshift range.\footnote{The narrow \SII\ ratios reported by \citet{NMFS19} are higher than those derived from our two component fitting because they include galaxies across the full redshift range (\mbox{0.6 $\lesssim z \lesssim$ 2.6}) whereas we focus only on galaxies at \mbox{1.9 $\lesssim z \lesssim$ 2.6}.} Together, these results suggest that the electron densities measured from the stacks of no-outflow inactive galaxies are likely to reflect the average conditions in \HII\ regions across the population of SFGs with \mbox{$\log(M_*/M_\odot) \gtrsim$ 9 -- 9.5} at the probed redshifts.

\subsection{Star Formation Driven Outflows}\label{appendix:outflow_densities}
Recent observational results suggest that star formation driven ionized gas outflows at high redshift are approximately five times denser than the ionized ISM in the galaxies from which the outflows are launched \citep{NMFS19}. The line emission from galaxies hosting outflows traces a combination of \HII\ region gas and outflowing material, and therefore the densities measured for these galaxies may be artificially enhanced and not representative of the conditions in \HII\ regions. We compare the densities measured from \textit{single} component emission line fits to stacked spectra of inactive galaxies with and without outflows (rather than the two component fitting method described in Appendix \ref{appendix:twocomp_fitting}), and test how including galaxies hosting outflows in our stacks impacts the average density measured for each redshift slice.

The \ktd+ parent sample includes 87 inactive galaxies hosting outflows, of which 82 have spectra covering the \SII\ doublet and 42 pass the quality control cut. 8/42 fall in the $z\sim$~0.9 redshift slice, 3/42 fall in the $z\sim$~1.5 slice, and 31/42 fall in the $z\sim$~2.2 slice. Due the relatively large sample variance (see e.g. Table \ref{table:filling_factor}), at least $\sim$~10 galaxies are required to obtain a meaningful measurement of the average \SII\ ratio, which in this case is only possible for the $z\sim$~2.2 subsample. Table \ref{table:k3d_yjhk_measurements} compares the pressures and densities measured for different subsamples of galaxies in each redshift slice. At $z\sim$~2.2, we measure somewhat higher densities and pressures in inactive galaxies with outflows compared to those with no outflows, although the two sets of measurements are consistent within the errors.\footnote{We note that the density measured from the outflow stack is $\sim$~49~\cmcubed\ higher than the density measured from the no outflow stack, but the ISM pressures derived from the two stacks differ by only 0.02 dex. This is because the outflow stack has a higher \NIIHa\ ratio, which results in a higher inferred metallicity and a lower inferred electron temperature compared to the no outflow stack. In reality, the enhanced \NIIHa\ ratio in the outflow stack is likely due to a small contribution from shock excitation rather than a higher metallicity \citep[e.g.][]{Davies19, Freeman19}, but this observation highlights the importance of accounting for metallicity differences between galaxy populations when studying ISM pressure.}

We also investigate how the inclusion of inactive galaxies hosting outflows in the primary sample would change the average measured pressure and density at each redshift. The fraction of inactive galaxies with outflows in our \SII-clean sample is similar to the fraction in the \ktd+ parent sample. Galaxies hosting outflows account for 8/47 or 17\% of inactive galaxies in our $z\sim$~0.9 slice compared to 14\% in the parent sample, 3/39 or 8\% in our $z\sim$~1.5 slice compared to 13\% in the parent sample, and 31/96 or 32\% in our $z\sim$~2.2 slice compared to 29\% in the parent sample. Therefore, the trends observed in our \SII-clean sample should reflect the trends in the overall inactive galaxy population at $z\sim$~0.9 and $z\sim$~2.2, but may slightly under-estimate the impact of outflows at $z\sim$~1.5. The measurements shown in Table \ref{table:k3d_yjhk_measurements} indicate that including inactive galaxies hosting outflows in the primary sample leads to modest enhancements in the inferred electron density (of up to $\sim$~35~\cmcubed) and ISM pressure (up to 0.2~dex), but the values derived from the full inactive sample and the no outflow subsample are consistent within the errors. These results confirm that the increased incidence of star formation driven outflows at \highz\ is unlikely to have a significant impact on the magnitude of the density evolution inferred from single component Gaussian fits to the \SII\ doublet lines.

The relatively small impact of outflows on the measured average densities can be explained by the modest fraction of inactive galaxies with detectable broad velocity components in their nebular emission line profiles. The \SII\ emission from the galaxies hosting outflows is a combination of an ISM component and an outflow component. For the $z\sim$~2.2 outflow galaxies, we measure a typical \SII\ ratio of 1.16. Assuming that the outflow component has a typical \SII\ ratio of 1.07 \citep[as measured by][]{NMFS19}, and that the ISM component has a typical \SII\ ratio of 1.22 (as measured from our stack of no-outflow inactive galaxies), the measured \SII\ ratio of 1.16 implies that $\sim$~40\% of the \SII\ flux originates from outflowing material. If we take this 40\% outflow flux fraction and multiply it by the fraction of inactive galaxies with outflows (32\% in our $z\sim$~2.2 slice), we would expect 13\% of the \SII\ emission from the full inactive galaxy population at $z\sim$~2.2 to be associated with outflowing material, corresponding to an expected \SII\ ratio of $\sim$~1.20 -- in very good agreement with the value measured from the `all inactive' stack. Performing a similar exercise for the $z\sim$~1.5 and $z\sim$~0.9 slices predicts that the average \SII\ ratios in the overall inactive galaxy population should 1.33 and 1.30, respectively, again in good agreement with the measured values.

Overall, we find that the presence of high density outflowing material enhances the densities measured for galaxies hosting outflows over those measured for galaxies without outflows, but the incidence of outflows in the inactive galaxy population is not sufficient to have a significant impact on the measured average densities. The outflow fraction is highest at $z\sim$~2.2 ($\sim$30\%), but at this redshift the typical ISM density is only a factor of $\sim$~2 lower than the typical density of the outflowing material, with the consequence that including sources with outflows still has a limited impact on the measured average density.

\begin{table}
\begin{nscenter}
\caption{Measured \SIIb/\SIIr, \SIIHa\ and \NIIHa\ ratios and calculated metallicities, electron densities and thermal pressures for different subsamples of \ktd+ galaxies in each redshift slice.}\label{table:k3d_yjhk_measurements}
\begin{tabular}{lccccccc}
\hline
Subsample & $n_{\rm gal}$ & \rSII\ & \NIIHa\ & \SIIHa\ & 12 + $\log$(O/H) & $\log(P/k)$ & $n_e$(\SII) (\cmcubed) \\ \hline \hline
\multicolumn{8}{c}{$z\sim$~0.9} \\ \hline
\textbf{Inactive (no outflow)} & 39 & 1.32$^{+0.11}_{-0.08}$ & 0.19 & 0.23 & 8.50 & 6.29$^{+0.28}_{-0.73}$ & 101$^{+59 }_{-85}$ \\
Inactive (with outflow) & 8 & \ldots & \ldots & \ldots \\
All inactive & 47 & 1.27$^{+0.10}_{-0.06}$ & 0.18 & 0.22 & 8.47 & 6.48$^{+0.14}_{-0.43}$ & 132$^{+60 }_{-83}$ \\
AGN hosts & 3 & \ldots & \ldots & \ldots \\
Inactive (no outflow) + AGN & 42 & 1.28$^{+0.09}_{-0.09}$ & 0.20 & 0.24 & 8.51 & 6.42$^{+0.24}_{-0.39}$ & 123$^{+87 }_{-77}$ \\  \hline
\multicolumn{8}{c}{$z\sim$~1.5} \\ \hline
\textbf{Inactive (no outflow)} & 36 & 1.34$^{+0.05}_{-0.13}$ & 0.16 & 0.24 & 8.39 & 6.23$^{+0.43}_{-0.26}$ & 79$^{+120 }_{-40}$ \\
Inactive (with outflow) & 3 & \ldots & \ldots & \ldots \\
All inactive & 39 & 1.33$^{+0.05}_{-0.13}$ & 0.16 & 0.24 & 8.39 & 6.27$^{+0.41}_{-0.22}$ & 89$^{+120 }_{-43}$ \\
AGN hosts & 16 & 1.18$^{+0.11}_{-0.09 }$ & \ldots & \ldots \\
Inactive (no outflow) + AGN & 52 & 1.25$^{+0.07}_{-0.08}$ & 0.17 & 0.27 & 8.37 & 6.58$^{+0.16}_{-0.25}$ & 157$^{+92 }_{-53}$ \\  \hline
\multicolumn{8}{c}{$z\sim$~2.2} \\ \hline
\textbf{Inactive (no outflow)} & 65 & 1.22$^{+0.15}_{-0.10}$ & 0.15 & 0.19 & 8.47 & 6.62$^{+0.20}_{-0.53}$ & 187$^{+140 }_{-132}$ \\
Inactive (with outflow) & 31 & 1.16$^{+0.17}_{-0.09}$ & 0.22 & 0.18 & 8.68 & 6.64$^{+0.16}_{-0.52}$ & 236$^{+130 }_{-160}$ \\
All inactive & 96 & 1.19$^{+0.12}_{-0.07}$ & 0.18 & 0.19 & 8.56 & 6.63$^{+0.14}_{-0.37}$ & 207$^{+104 }_{-105}$ \\
AGN hosts & 14 & 1.11$^{+0.26}_{-0.19 }$ & \ldots & \ldots \\
Inactive (no outflow) + AGN & 79 & 1.22$^{+0.13}_{-0.10}$ & 0.16 & 0.20 & 8.46 & 6.61$^{+0.17}_{-0.55}$ & 182$^{+142 }_{-122}$ \\ \hline
\end{tabular}
\end{nscenter}
\tablecomments{In each section of the table, the top row gives the `fiducial' ISM pressure and electron density measured from the stack of inactive galaxies with no outflows. Subsequent rows list measurements for inactive galaxies with outflows and AGN host galaxies, and show how the density and pressure measurements change when these subpopulations are included in the primary sample. We did not make measurements for stacks with less than ten galaxies due to the relatively large sample variance. Densities and pressures are not calculated for AGN hosts because the \HII\ region model grids that are used to convert from \rSII\ to density and pressure cannot be meaningfully applied to spectra excited by non-stellar sources.}
\end{table}

\subsection{AGN Contamination}\label{appendix:agn_densities}
There is growing observational evidence to suggest that AGN-driven outflows have \SII\ densities of $\gtrsim$~1000~\cmcubed\ \citep[e.g.][]{Perna17, Kakkad18, Husemann19, NMFS19, Shimizu19}; significantly denser than the ISM of typical inactive galaxies at \highz. Furthermore, outflows appear to be almost ubiquitous in AGN-host galaxies at $z\sim$~1~--~2 \citep[e.g.][]{Genzel14, NMFS14, Harrison16, NMFS19, Leung19}. We perform tests similar to those described in the previous section to investigate the impact of AGN contamination on measurements of the ISM pressure and electron density at \highz.

The \ktd+ parent sample contains 136 AGN host galaxies, of which 112 have spectra covering the \SII\ doublet. One of the galaxies is classified as a Type 1 AGN and excluded, and 33 of the remaining 111 galaxies pass the quality control cut. 3/33 fall in the $z\sim$~0.9 redshift slice, 16/33 fall in the $z\sim$~1.5 slice, and 14/33 fall in the $z\sim$~2.2 slice. We note that only 6/33 are classified as no-outflow systems (consistent with the high outflow fraction), so we analyze all of the AGN hosts together. AGN hosts account for 3/50 or 6\% of galaxies in our $z\sim$~0.9 slice compared to 23\% in the parent sample, 16/55 or 29\% in our $z\sim$~1.5 slice compared to 29\% in the parent sample, and 14/110 or 13\% in our $z\sim$~2.2 slice compared to 20\% in the parent sample. The AGN fraction in our \SII-clean sample is very similar to the fraction in the parent sample for the $z\sim$~1.5 slice, but is significantly lower than the fraction in the parent sample for the other two redshift slices, indicating that our tests may underestimate the true impact of AGN contamination in these redshift ranges. 

The \SII\ ratios measured for the AGN host galaxies in the $z\sim$~1.5 and $z\sim$~2.2 slices are listed in Table \ref{table:k3d_yjhk_measurements}. We do not present density or pressure measurements for the AGN host galaxies because the grids that are used to convert the \SII\ ratios to densities and pressures are outputs of \HII\ region photoionization models. The hard ionizing radiation field of the AGN will change the ionization and temperature structure of the nebula, resulting in a different relationship between \rSII\ and density/pressure. AGN host galaxies appear to have lower \SII\ ratios than inactive galaxies with and without outflows at the same redshift, although the differences between the line ratios measured for the different $z\sim$~2.2 stacks are not statistically significant. There is some evidence to suggest that AGN contamination may impact the measured ISM densities and pressures, particularly in the $z\sim$~1.5 slice where both quantities increase by a factor of $\sim$~2 when AGN are included. The inclusion of AGN host galaxies has only a minor impact on measured average density in the $z\sim$~0.9 redshift slice, and does not have any significant impact on the measurements in the $z\sim$~2.2 slice, but we emphasise that for these redshift slices, the AGN fractions in our \SII-clean sample are a factor of $\sim$~2~--~4 lower than in the parent sample, and therefore our test likely only provides a lower limit on the impact of AGN contamination. 

In studies of high-redshift star-forming galaxies, AGN are commonly identified using \mbox{X-ray}, radio and/or mid-IR data. However, \citet{NMFS19} showed that these classical selection techniques may miss up to $\sim$~50\% of AGN at $z\sim$~1~--~2, likely due to both the varying availability and depth of ancillary data between different extragalactic fields, and the impact of AGN variability and small-scale nuclear obscuration on the different diagnostic tracers \citep[e.g][]{Padovani17}. Our results indicate that it is important to use conservative AGN selection criteria (e.g. including optical line ratios) to minimize the probability of contamination from non-stellar excitation sources when measuring electron densities and ISM pressures.

\section{High-$z$ Literature Samples}\label{sec:literature_samples}
We supplement our \ktd+ results with measurements from other surveys of \highz\ galaxies in the literature. Specifically, we select samples of galaxies at $z \gtrsim$~0.6 that lie within $\sim$~0.5~dex of the \citet{Speagle14} star-forming MS and have published \SIIb/\SIIr\ or \OII$\lambda$3726/\OII$\lambda$3729 measurements. All stellar masses and SFRs are scaled to the \citet{Chabrier03} initial mass function for consistency with our measurements. We require published line ratios to avoid systematic biases in the conversion between line ratios and densities arising from differences in the atomic data or assumed electron temperature (see e.g. discussions in \citealt{Sanders16}, \citealt{Ke19_density}). We only include datasets with at least 20 galaxies to minimize the impact of variance within the galaxy population. Our final \highz\ comparison sample contains measurements from KROSS \citep{Swinbank19}, COSMOS-\OII\ \citep{Kaasinen17}, FMOS-COSMOS \citep{Kashino17}, MOSDEF \citep{Sanders16}, and KBSS-MOSFIRE \citep{Steidel14}. Table \ref{table:literature} summarizes the properties of the \highz\ literature samples as well as the line ratio measurements and the derived electron densities. We do not have size measurements for the galaxies in these samples, so the average surface density quantities used in Figures \ref{fig:rsii_sf} and \ref{fig:rsii_stars_gas} are estimated assuming that the galaxies lie on the redshift-dependent mass-size relation from \citet{vanderWel14}. 

\begin{table*}
\begin{nscenter}
\caption{Literature Samples}\label{table:literature}
\begin{tabular}{lcccccc}
\hline Sample/Reference & Median & Median & Median & Density & Line Ratio & $n_e$ (\cmcubed) \\ 
 & $z$ &  $\log(M_*/M_\odot)$ & SFR ($M_\odot$ yr$^{-1}$) & Diagnostic & & \\ \hline
KROSS \citep{Swinbank19} & 0.85 & 10.0 & 6.7 & \SII\ & 1.4~$\pm$~0.1 & 33$^{+78}_{-32}$ \\
COSMOS-\OII\ \citep{Kaasinen17} & 1.5 & 10.6 & 26.4 & \OII\ & 1.29~$\pm$~0.03 & 119$_{-21}^{+24}$ \\ 
FMOS-COSMOS \citep{Kashino17} & 1.55 & 10.2 & $\sim$~59$^a$ & \SII\ & 1.21~$\pm$~0.1 & 193$_{-88}^{+150}$ \\ 
MOSDEF \citep{Sanders16} & 2.24 & 10.1 & 29.7 & \SII\ & 1.13$^{+0.16}_{-0.06}$ & 316$^{+92}_{-197}$ \\ 
KBSS-MOSFIRE \citep{Steidel14} & 2.3 & 10.0 & 23.3 & \OII\ & 1.16~$\pm$~0.04 & 258$_{-56}^{+67}$ \\ \hline
\end{tabular}
\end{nscenter}
\tablecomments{Line ratio measurements were taken directly from the listed references and electron densities were re-calculated adopting the metallicity of the \ktd\ stack lying closest in redshift. In cases where both the \SII\ and the \OII\ ratio were quoted we calculate the density from the \SII\ ratio for consistency with our analysis. $^a$From Figure 1 of \citet{Kashino17} we estimated that their \Ha-detected sample has a median SFR of $\sim$~100~$M_\odot$~yr$^{-1}$. Their SFRs were calculated assuming a Salpeter IMF, so we divided by 1.7 to convert to a Chabrier IMF, yielding the $\sim$~59~$M_\odot$~yr$^{-1}$ quoted here.}
\end{table*}

\section{Constraints on Atomic Gas Reservoirs at $z\sim$~1-3}\label{appendix:HI}
The redshift evolution of galaxy atomic gas reservoirs is poorly constrained because current radio telescopes can only detect \HI\ emission from galaxies at \mbox{$z\lesssim$~0.4}. At higher redshifts, the atomic gas mass volume density can be estimated from Ly$\alpha$ absorption in quasar spectra, but this probes gas both in galaxies and in the circumgalactic medium around galaxies. Current observational compilations suggest that the neutral hydrogen volume density of the universe has decreased by a factor of $\sim$~1.5 since $z\sim$~2 (see \citealt{Peroux20} and references therein), whereas the molecular hydrogen volume density has decreased by a factor of $\sim$~4 \citep[e.g.][]{Decarli16, Scoville17, Riechers19, Lenkic20, Tacconi20}. Therefore, the fraction of the ISM in the molecular phase is expected to increase towards higher redshifts.

\HI\ mapping surveys of local spiral galaxies have found that $\Sigma_{HI}$ has a relatively constant value of $\sim$~6~$M_\odot$~pc$^{-2}$ at all galactocentric radii \citep[e.g.][]{Leroy08} and rarely exceeds $\sim$~10~$M_\odot$ pc$^{-2}$ \citep[e.g.][]{Bigiel08, Leroy08, Schruba18}, most likely because \HI\ is converted to $H_2$ at higher gas mass surface densities (see e.g. discussion in \citealt{Tacconi20}). The maximal $\Sigma_{HI}$ is inversely correlated with the gas phase metallicity \citep[e.g.][]{Schruba18}. The metallicities measured from our four stacked spectra (see Section \ref{subsubsec:measurements}) are consistent with each other within 0.1 dex, because the average stellar mass of the probed galaxies increases towards higher redshifts. Therefore, the maximal $\Sigma_{HI}$ is not expected to vary significantly between our samples. On the other hand, \gmsd\ increases from the outskirts to the centers of local spiral galaxies and no saturation is observed \citep[e.g.][]{Schruba11}.

There is a tight relationship between the masses and diameters of local \HI\ disks which implies a uniform characteristic $\Sigma_{HI}$ of 6.9~$M_\odot$~pc$^{-2}$ \citep{Broeils97, Wang16} (assuming a standard helium mass fraction of 36\%). In comparison, the \citet{Tacconi20} molecular gas depletion time scaling relation suggests that the SAMI and \ktd+ galaxies have median molecular gas mass surface densities of \mbox{$\sim$~8~$M_\odot$~pc$^{-2}$} and \mbox{$\sim$~200~$M_\odot$~pc$^{-2}$}, respectively (see Figure \ref{fig:rsii_stars_gas}). The typical \gmsd\ for the \ktd+ galaxies is more than an order of magnitude above the surface density at which $\Sigma_{HI}$ is observed to saturate in local galaxies, suggesting that the ISM in the central regions of \highz\ SFGs is likely to be strongly dominated by molecular gas. On the other hand, the gas reservoirs within $R_e$ for the SAMI galaxies are likely to be approximately equal parts \HI\ and $H_2$. We adopt \mbox{$\Sigma_{HI}$~=~6.9~$M_\odot$~pc$^{-2}$} at all redshifts, but note that the average total gas mass surface densities of the \ktd+ galaxies are insensitive to the adopted $\Sigma_{HI}$ within any reasonable range of values. Varying $\Sigma_{HI}$ by a factor of 2 would change the median midplane pressure at $z\sim$~0 (Section \ref{subsec:ambient_pressure}) by $\lesssim$~0.2 dex and the inverse of the \HII\ region stall radius (Section \ref{subsec:stall_radius}) by $\lesssim$~0.15 dex.

\section{Stellar Velocity Dispersion Estimates}\label{appendix:sigma_stellar}
For a disk in hydrostatic equilibrium, the vertical stellar velocity dispersion is given by:
\begin{equation}
\sigma_{*,z} = h_* \sqrt{2 \pi G  \rho} 
\end{equation}
where $h_*$ is the scale height of the stellar disk and $\rho$ is the midplane mass volume density \citep{vanderKruit88, Leroy08}. We assume that, for the relatively high mass galaxies considered in this analysis, $\rho$ is dominated by the stellar and gas components.

In the local universe, the average `flattening' (ratio of scale length to scale height, $R_{d,*}/h_*$) of the stellar disk is 7.3 \citep{Kregel02, Sun20}. For the SAMI galaxies at $z\sim$~0 we adopt \mbox{$h_*$ = $R_{d,*}$/7.3}, and approximate $R_{d,*}$ from $R_e$ using the relationship for an exponential disk (\mbox{$R_{d,*} \simeq$ $R_e$/1.67}).

High-$z$ disks are significantly thicker than $z\sim$~0 disks, both photometrically \citep[e.g.][]{Elmegreen17} and kinematically \citep[e.g.][]{NMFS09, Wisnioski15, Johnson18}, and therefore we cannot use the $z\sim$~0 flattening to derive $h_*$ from $R_e$. Instead, we assume that the stellar scale heights of the \ktd+ galaxies are approximately equivalent to their ionized gas scale heights. This is plausibly a reasonable assumption (to first order) given that the majority of the ionized gas is likely to be associated with \HII\ regions (see discussion in Section \ref{subsubsec:dig_contribution}) and that the stellar populations of galaxies at $z\sim$~1~--~2 must be relatively young. The scale heights of stellar disks increase over time as a result of mergers and gravitational interactions with massive objects in the disk (such as giant molecular clouds, globular clusters and black holes). These encounters randomly perturb the momentum of individual stars, leading to the diffusion of stellar orbits and an increase in the vertical velocity dispersion and scale height of the stellar disk \citep[e.g.][]{Wielen77}. The scale height of the star-forming disk also changes over time (see discussion in Section \ref{subsubsec:scale_heights}). Therefore, the stellar and ionized gas scale heights will be most similar in galaxies with young stellar populations. We note that varying $h_*$ by a factor of 2 changes the midplane pressures derived in Section \ref{subsec:ambient_pressure} by $<$~0.15~dex and therefore the uncertainties on $h_*$ have a relatively limited impact on our results.

We have a single value of $\rho$ per galaxy which can be used to estimate the average stellar velocity dispersion within $R_e$. We assume that the scale height is constant as a function of radius (for consistency with the $n_e$(rms) calculations in Section \ref{subsec:rms_density}), but note that the average scale height within $R_e$ for a flared disk (i.e. with constant disk velocity dispersion; e.g. \citealt{Burkert10}) is almost identical to the scale height at $R_d$, and therefore the choice of geometry has a negligible impact on the derived midplane pressure. 

\section{Derivation of \HII\ Region Stall Radius}\label{appendix:rstall}
The time evolution of the internal pressure of an adiabatically expanding bubble driven by stellar winds and supernovae is given in Equation 25 of \citet{Oey97}:
\begin{equation}
P_{\rm int}(t) = \frac{7}{(3850\pi)^{2/5}}~L_{\rm mech}^{2/5}~\rho_0^{3/5}~t^{-4/5}
\label{eqn:p_int_t}
\end{equation}
where $L_{\rm mech}$ is the mechanical luminosity of the central star cluster and $\rho_0$ is the mass volume density of the material surrounding the \HII\ region. Assuming that the expansion of the bubble stalls when $P_{\rm int}$ reaches the ambient pressure $P_0$, the stall time $t_{\rm stall}$ is related to $L_{\rm mech}$, $\rho_0$ and $P_0$ as follows:
\begin{equation}
\begin{split}
t_{\rm stall} &\propto L_{\rm mech}^{1/2}~\rho_0^{3/4}~P_0^{-5/4}
\label{eqn:t_stall}
\end{split}
\end{equation}
The time evolution of the radius of the expanding bubble is given in Equation 24 of \citet{Oey97}:
\begin{equation}
R(t) = \left( \frac{250}{308\pi} \right)^{1/5}~L_{\rm mech}^{1/5}~\rho_0^{-1/5}~t^{3/5}
\label{eqn:r_t}
\end{equation}
The expression for the stall time (Equation \ref{eqn:t_stall}) can be substituted into Equation \ref{eqn:r_t} to derive the stall radius $R_{\rm stall}$ as a function of $L_{\rm mech}$, $\rho_0$ and $P_0$:
\begin{equation}
 \begin{split}
R_{\rm stall} &\propto L_{\rm mech}^{1/5}~\rho_0^{-1/5}~\left(L_{\rm mech}^{1/2}~\rho_0^{3/4}~P_0^{-5/4} \right)^{3/5} \\
&\propto L_{\rm mech}^{1/2}~\rho_0^{1/4}~P_0^{-3/4} \\
 \end{split}
\end{equation}
We assume that the ambient pressure is equivalent to the hydrostatic equilibrium midplane pressure $P_{\rm mid}$ (discussed in Section \ref{subsec:ambient_pressure}), that the mechanical luminosity of the central star cluster is proportional to \sfrsd\ (discussed in Section \ref{subsec:feedback_pressure}), and that the mass volume density of the material surrounding the \HII\ region is proportional to $\rho_{H_2}$ (discussed in Section \ref{subsec:birth_cloud_density}). Under these conditions, the stall radius can be expressed as
\begin{equation}
R_{\rm stall} \propto \Sigma_{\rm SFR}^{1/2}~\rho_{H_2}^{1/4}~P_{\rm mid}^{-3/4} \\
\label{eqn:rstall}
\end{equation}

\bibliography{../bibliography/mybib}

\end{document}